\documentclass[journal,draftcls,onecolumn,12pt,twoside]{IEEE}
\usepackage{multirow}
\usepackage{amssymb}
\usepackage[dvips]{graphicx}
\usepackage{amsmath}
\usepackage{amsthm}
\usepackage{latexsym,bm}
\usepackage{color}
\usepackage{subfigure}
\usepackage{longtable}
\usepackage{accents}
\usepackage{cite}
\usepackage{enumerate}
\usepackage{arydshln}
\usepackage{setspace}
\usepackage{booktabs}
\usepackage{array, caption, threeparttable}

\usepackage{algorithmic}
\usepackage{algorithm}

\usepackage{diagbox}
\makeatletter
\def\widebar{\accentset{{\cc@style\underline{\mskip10mu}}}}
\def\Widebar{\accentset{{\cc@style\underline{\mskip8mu}}}}
\makeatother

\allowdisplaybreaks
\theoremstyle{plain}

\theoremstyle{definition}
\theoremstyle{definition} 
\renewcommand{\IEEEQED}{\IEEEQEDopen}
\begin{document}

\title{Performance Analysis and Resource Allocation of STAR-RIS Aided Wireless-Powered NOMA System
\thanks{K. Xie and G. Cai are with the School of Information Engineering, Guangdong University of Technology, China (e-mail: xiekengyuan@126.com, caiguofa2006@gdut.edu.cn).}
\thanks{G. Kaddoum is with the University of Qu$\acute{\mathrm{e}}$bec, Qu$\acute{\mathrm{e}}$bec, QCG1K 9H7, Canada, and also with $\acute{\mathrm{E}}$cole de Technologie Sup$\acute{\mathrm{e}}$rieure ($\acute{\mathrm{E}}$TS), LaCIME Laboratory, Montreal, QC H3C 1K3, Canada (e-mail: georges.kaddoum@etsmtl.ca).}
\thanks{J. He is with the Technology Innovation Institute, 9639 Masdar City, Abu Dhabi, United Arab Emirates (E-mail: jiguang.he@tii.ae).}
}
\author{Kengyuan Xie, Guofa Cai, Georges Kaddoum, {\em Senior Member, IEEE}, Jiguang He, \IEEEmembership{Senior Member, IEEE} }

\maketitle
\vspace{-1.6cm}
\begin{abstract}
This paper proposes a simultaneous transmitting and
reflecting reconfigurable intelligent surface (STAR-RIS) aided wireless-powered non-orthogonal multiple access (NOMA) system, which includes an access point (AP), a STAR-RIS, and two non-orthogonal users located at both sides of the STAR-RIS.
In this system, the users first harvest the radio-frequency energy from the AP in the downlink, then adopt the harvested energy to transmit information to the AP in the uplink concurrently.
Two policies are considered for the proposed system. The first one assumes that the time-switching protocol is used in the
downlink while the energy-splitting protocol is adopted in the
uplink, named TEP. The second one assumes that the energy-splitting protocol is utilized in both the downlink and uplink, named EEP.
The outage probability, sum throughput, and average age of information (AoI) of the proposed system with TEP and EEP are investigated over Nakagami-$m$ fading channels.
In addition, we also analyze the outage probability, sum throughput, and average AoI of the STAR-RIS aided wireless-powered  time-division-multiple-access (TDMA) system.
Simulation and numerical results show that the proposed system with TEP and EEP outperforms baseline schemes, and significantly improves sum throughput performance but reduces outage probability and average AoI performance compared to the STAR-RIS aided wireless-powered TDMA system.
Furthermore, to maximize the sum throughput and ensure a certain average AoI,
we design a genetic-algorithm based time allocation and power allocation (GA-TAPA) algorithm.
Simulation results demonstrate that the proposed GA-TAPA method can significantly improve the sum throughput by adaptively adjusting system parameters.
\vspace{-0.5cm}
\end{abstract}
\vspace{-0.5cm}
\section{Introduction} \label{sect:review}
Trillions of Internet-of-Things (IoT) devices will emerge, especially with the explosive growth of low-power devices, which will require a rethinking of future network design\cite{9502719}.
For the design of reliable and robust networks, the challenge of energy-constrained IoT devices is energy supply\cite{9128010}.
Recently, energy harvesting (EH) has been proposed as one of the most promising candidates to solve the energy-supply problem\cite{9333647}.
For example, harvesting energy from solar, piezoelectric, electromagnetic and other sources was proven to be effective for powering IoT devices\cite{9383098}.
In particular, radio frequency (RF)
based EH provides an attractive solution to power low-power IoT devices over the air due to its flexible and reliable characteristics\cite{8843917}.
As a result, it enables wireless-powered communication (WPC)\cite{9615376},
which combines both wireless information transfer (WIT) and wireless power transfer (WPT)\cite{8982086}. WPC has advantages in reducing the operational cost compared to conventional battery-powered counterparts and in improving the robustness of wireless communication networks especially low-power sensor networks. The major challenge of the WPC is the low transfer efficiency over a long distance\cite{9298890}.
Although the performance of WPC systems can be improved by adopting various existing techniques, e.g., relaying\cite{8648303} and multiple-input multiple-output (MIMO)\cite{8955833}.
However, high energy consumption and hardware cost are introduced by these techniques due to signal amplification or regeneration and
a large number of RF chains.

Recently, reconfigurable intelligent surfaces (RISs), which can enhance spectrum efficiency, energy efficiency, and physical-layer
security\cite{9398559}, have been proposed as one of the key technologies for the sixth-generation (6G) wireless networks\cite{9558795,9326394}.
RIS, which consists of a large number of low-cost reflective elements, is an economical and energy-efficient technology compared to MIMO and relaying systems\cite{9326394}.
To improve the WPT efficiency and data rates of the WPC system, there has been  a great interest in RIS aided WPC \cite{9003222,9409104,10011440,9408423}.
In \cite{9003222}, a RIS-assisted cooperative WPC network (WPCN) was proposed, where the
RIS is utilized to improve the energy efficiency of the WPT phase and the spectrum efficiency of the WIT phase.
In \cite{9409104}, a RIS aided wireless-powered sensor network was studied, where the
RIS is deployed to enhance the sum throughput by intelligently adjusting the phase shift of each reflecting element.
In \cite{10011440}, an energy buffer and RIS aided WPC system was proposed to improve the error performance.
In \cite{9408423}, it was revealed that the doubly near-far issue in MIMO WPCN can be solved effectively with a rigorous deployment of the RIS.

However, the conventional RIS requires both the access point (AP) and user to be on the same side of the RIS\cite{9739715}.
To overcome this drawback, the concept of simultaneous transmitting and reflecting RIS (STAR-RIS) was proposed in \cite{9437234,9690478}.
Different from conventional RIS, each element of STAR-RIS can transmit and reflect the incident signal simultaneously, thus breaking the location limitation of RIS deployment and achieving \textit{full-space coverage}\cite{9740451}.
In STAR-RIS, one part of the incident signal is reflected to the same space as the incident signal, i.e., the reflection space, while the other part of the incident signal is transmitted to the opposite space, i.e., the transmission space\cite{9570143}.
Moreover, STAR-RIS provides a new degree-of-freedom for manipulating signal propagation, which increases the flexibility for network design\cite{9462949}.
To exploit the benefits of both WPC and STAR-RIS, a STAR-RIS-enhanced wireless-powered mobile edge computing system was proposed in \cite{111111}, which was shown to improve the efficiency of energy transfer and task offloading by maximizing the total computation rate of all users.
Furthermore, three practical protocols, including energy-splitting, mode-switching, and time-switching, were presented in \cite{9570143}.
In the energy-splitting and mode-switching protocols, since the STAR-RIS splits the incident signal into two parts, a multiple access scheme need to be designed to distinguish these two parts for successful decoding\cite{9808307}.
Existing multiple access schemes feature two categories: orthogonal multiple access (OMA) and non-orthogonal multiple access (NOMA)\cite{9740451}.
Compared to OMA, NOMA can provide better spectrum efficiency and user fairness\cite{6868214,7676258}.
Recently, several research efforts have focused on the performance analysis of STAR-RIS aided NOMA networks \cite{9462949,9722712,9774334,9786807}.
In \cite{9462949}, the basic coverage of the STAR-RIS aided NOMA network was studied.
In \cite{9722712}, the outage probability and diversity gains of a STAR-RIS aided downlink NOMA network with randomly deployed users were derived.
In \cite{9774334}, the outage probability analysis of the STAR-RIS aided NOMA network over correlated channels was considered.
In \cite{9786807}, the error performance of the STAR-RIS aided NOMA network was analyzed.
In these existing STAR-RIS aided NOMA works\cite{9462949,9722712,9774334,9786807}, the STAR-RIS is only deployed in the downlink transmissions. Actually, the STAR-RIS can be adopted in the uplink transmissions to improve the performance. However, the analytical methods of downlink transmissions for the STAR-RIS aided NOMA system can not be directly applicable for the uplink transmission.

With the aforementioned motivations, in this paper, we propose a STAR-RIS aided wireless-powered NOMA system, where the STAR-RIS is utilized to improve the efficiency of the WPT in the downlink and the performance of the WIT in the uplink.
Moreover, we investigate the outage probability and sum
throughput performance of the proposed system over Nakagami-$m$
fading channels.
Furthermore, the information freshness is essential for the IoT networks \cite{9144239}. The age of information (AoI) is a metric for information freshness and plays a key role in real-time operations \cite{6195689}.
For this reason, we further analyze the average AoI of the proposed system.
The contributions of this paper are summarized as follows:
\begin{itemize}
\item
A STAR-RIS aided wireless-powered NOMA system is put forward, where the optimal decoding order policy for the successive interference cancellation (SIC) is considered instead of following a specific decoding order.
Moreover, two policies, i.e., TEP and EEP, are designed for the proposed system.
Specifically, for the TEP,  the time-switching and energy-splitting protocols are adopted in the
downlink and uplink, respectively.
For the EEP, the energy-splitting protocol is used in both the downlink and uplink.
\item
Based on the moment-matching approach, the outage probability and sum throughput expressions for the proposed system with TEP and EEP are derived over Nakagami-$m$ fading channels.
Moreover, to meet the timeliness requirements, the closed-form average AoI expressions of the proposed system with TEP and EEP is further investigated. In addition, the outage probability, sum throughput, and average AoI of the STAR-RIS aided wireless-powered time-division-multiple-access (TDMA) system are also analyzed.
Simulation and numerical results show that the proposed system with TEP and EEP offers improved performance compared to baseline schemes, and significantly improves sum throughput performance but reduces outage
probability and average AoI performance compared to the STAR-RIS aided wireless-powered TDMA system.

\item
To maximize the sum throughput and ensure a certain average AoI,
a genetic-algorithm based time allocation and power allocation (GA-TAPA) algorithm, which jointly optimizes the time allocation and the power allocation, is proposed.
Simulation results show that the proposed GA-TAPA
method can provide a higher sum throughput performance by adaptively adjusting the system parameters.
\end{itemize}
\vspace{-0.3cm}

The remainder of this paper is organized as follows.
Section II introduces the system model.
Section III analyzes the outage probability and sum throughput performance of the proposed system.
Section IV carries out the average AoI analysis.
Section V  presents  the proposed resource allocation scheme.
Section VI presents the numerical results and discussions while
Section VII concludes the paper.

\section{System Model} \label{sect:system model}
In this section, we describe the proposed STAR-RIS aided wireless-powered NOMA system. Two policies are considered for the proposed system, and their signal models are presented. Finally, we also introduce the STAR-RIS aided wireless-powered TDMA system.

\begin{figure}[!htb]
\begin{tabular}{cc}
\begin{minipage}[t]{0.48\linewidth}
    \includegraphics[width = 1\linewidth]{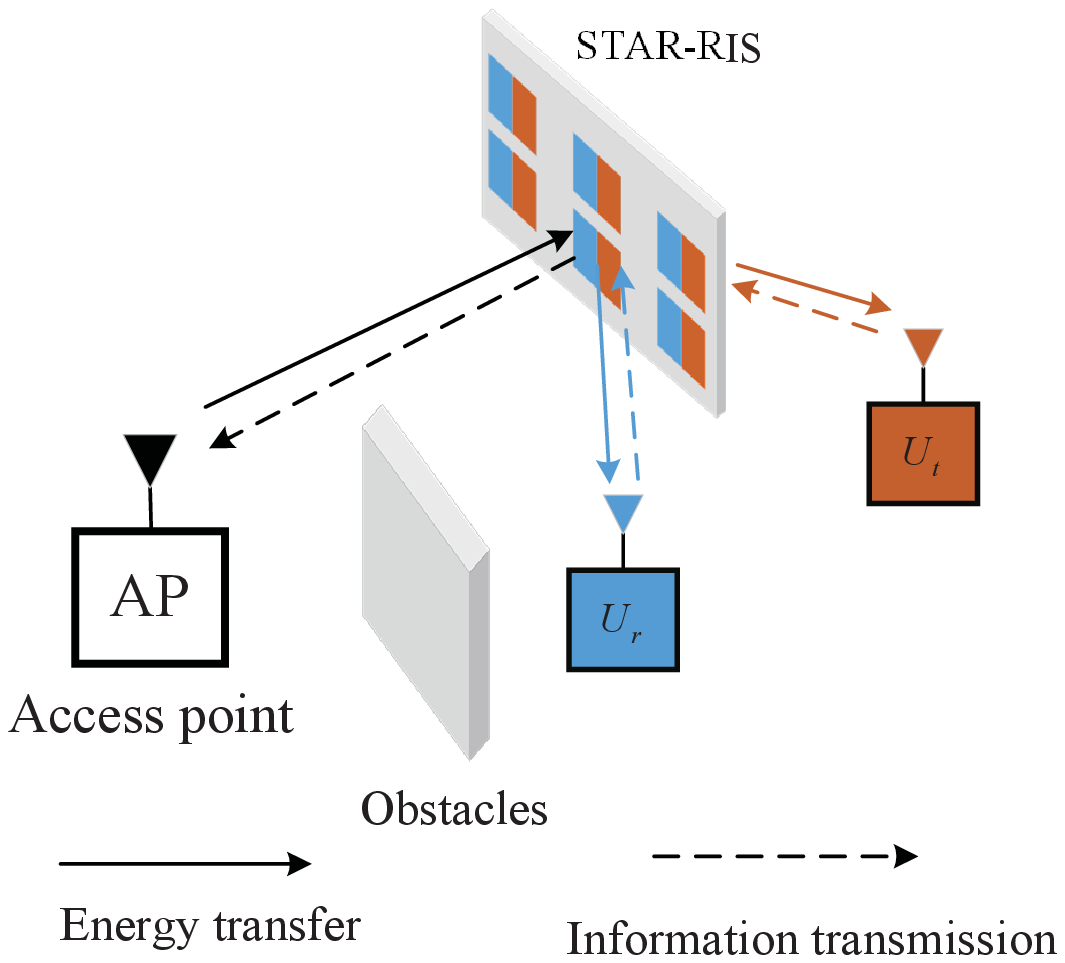}
    \vspace{-0.8cm}
    \captionsetup{font={small}, justification=raggedright}
    \caption{STAR-RIS aided wireless-powered NOMA system model.}
    \label{fig:system}
\end{minipage}
\begin{minipage}[t]{0.48\linewidth}
    \includegraphics[width = 1\linewidth]{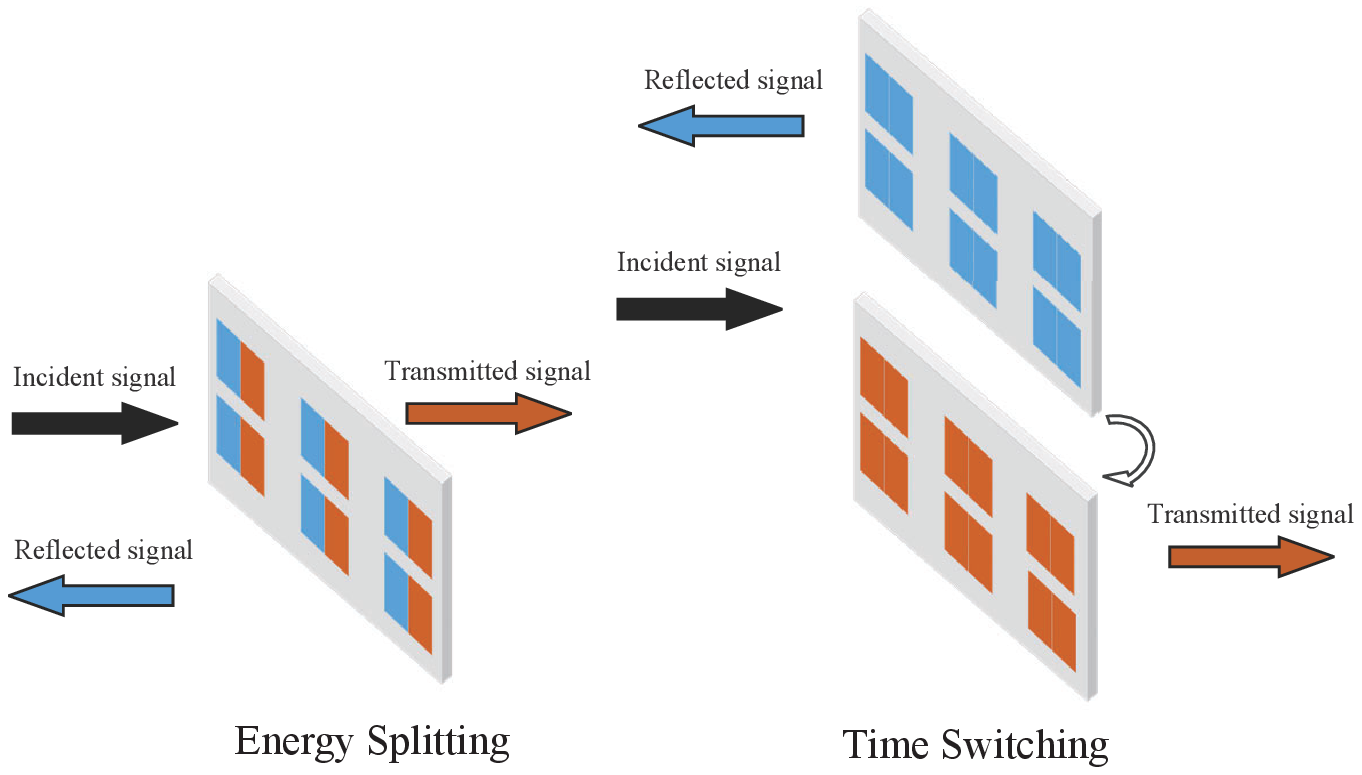}
    \vspace{-0.8cm}
    \captionsetup{font={small}, justification=raggedright}
    \caption{Energy-splitting and time-switching protocols.}
    \label{fig:model}
\end{minipage}
\end{tabular}
\vspace{-1.2cm}
\end{figure}

\vspace{-0.5cm}

\subsection{STAR-RIS Aided Wireless-Powered NOMA System}
A STAR-RIS aided wireless-powered NOMA system is shown in Fig.~\ref{fig:system}, which consists of an AP, a STAR-RIS, and two users ${U_r}$ and ${U_t}$.
Here, AP, ${U_r}$, and ${U_t}$ are equipped with a single antenna.
In this system, it is assumed that there is a fixed energy supply for the AP. ${U_r}$ and ${U_t}$ are energy-constrained nodes and have to harvest the energy from the RF signal of the AP in the downlink.
Then, in the uplink, ${U_r}$ and ${U_t}$ simultaneously transmit the backlogged data to the AP using the harvested energy in a NOMA fashion.
The STAR-RIS, composed of $N$  low-cost reflective elements,
assists energy transfer (ET) from the AP to ${U_\chi}$ in the downlink and information transmission (IT) from the ${U_\chi}$ to AP in the uplink, where $\chi  \in \left\{ {t,r} \right\}$.

The energy-splitting and time-switching protocols are considered, as shown in Fig.~\ref{fig:model}.
In TEP, the time-switching protocol is used in the uplink while the energy-splitting protocol is adopted in the downlink.
Meanwhile, in EEP, the energy-splitting protocol is utilized in both the uplink and downlink.

\begin{figure}
\center
\includegraphics[width=2.9in,height=1.4in]{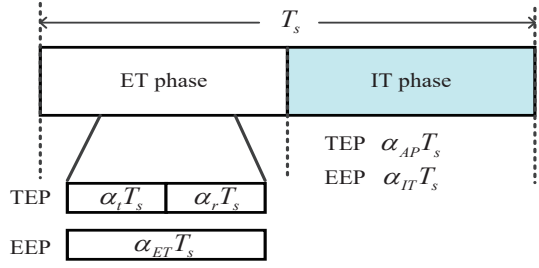}
\vspace{-0.2cm}
\captionsetup{font={small}}
\caption{ET and IT phases for the proposed system.}
\label{fig:time}
\vspace{-1.2cm}
\end{figure}

The ET and IT phases of the TEP and EEP schemes are shown  Fig.~\ref{fig:time}.
In the TEP scheme, the total communication duration $T_s$ is divided into three phases, where the ET phase for $U_t$ is ${\alpha _{t}}{T_s}$, the ET phase for $U_r$ is ${\alpha _{r}}{T_s}$, and the IT phase is ${\alpha _{AP}}{T_s}$, where the coefficients $\alpha_t$, $\alpha_r$, and $\alpha_{AP}$ satisfy $ 0< \alpha_t, \alpha_r, \alpha_{AP}<1$ and  ${\alpha _t} + {\alpha _r} + {\alpha _{AP}} = 1$.
In the EEP scheme, the duration of the ET phase is ${\alpha _{ET}}{T_s}$ and the duration of the IT phase is ${\alpha _{IT}}{T_s}$, where the coefficients $\alpha_{ET}$ and $\alpha_{IT}$ satisfy $ 0< \alpha_{ET}, \alpha_{IT}<1$ and  ${\alpha _{ET}} + {\alpha _{IT}} = 1$.
Without loss of generality, in this paper, we consider a normalized unit communication block time in the sequel, i.e., ${{T_s}} = 1$.
The signal models of the TEP and EEP schemes are presented in the following.
\vspace{-0.5cm}
\subsection{TEP}
\subsubsection{ET phase}
For the time-switching protocol, the STAR-RIS switches all elements between the transmission and reflection modes in different time periods.
The ET phase is divided into two periods, i.e., the transmission period of duration ${\alpha _t}{T_s}$ for ${U_t}$ and the reflection period of duration ${\alpha _r}{T_s}$ for ${U_r}$.
Let ${\boldsymbol{ \varphi} ^{TEP}_{ET}} \buildrel \Delta \over = \left[ {{e^{j{\varphi _1^{TEP}}}}, \ldots ,{e^{j{\varphi _N^{TEP}}}}} \right]$ and ${\boldsymbol{ \zeta}^{TEP}_{ET}} \buildrel \Delta \over = \left[ {{e^{j{\zeta _1^{TEP}}}}, \ldots ,{e^{j{\zeta _N^{TEP}}}}} \right]$ be the transmission and reflection coefficient vectors of the STAR-RIS, respectively, where $\varphi _n^{TEP}$ and $\zeta _n^{TEP} \in \left[ {0,2\pi } \right)$ are the adjustable phase-shifts on the $n$-th element, $n \in \left\{ {1,2, \ldots ,N} \right\}$, $j$ denotes the imaginary unit, i.e., ${j^2} =  - 1$.

Let ${\mathbf{\tilde {h}}} \in {\mathbb{C} ^{1 \times N}}$ denote the \textit{complex channel coefficients} between the AP and STAR-RIS, ${\mathbf{\tilde {g}_r}} \in {\mathbb{C}^{1 \times N}}$ the \textit{complex channel coefficients} between the STAR-RIS and ${U_r}$, and ${\mathbf{\tilde {g}_t}} \in {\mathbb{C}^{1 \times N}}$ the \textit{complex channel coefficients} between the STAR-RIS and ${U_t}$, where ${\mathbb{C}^{1 \times N}}$ is the space of $1 \times N$ complex-valued vectors.
The path loss coefficient is ${l_\chi } = \frac{1}{{d_0^{{\vartheta _0}}d_\chi ^{{\vartheta _\chi }}}},$
where ${d_0}$ and ${d_\chi}$ are the distances from the AP to the STAR-RIS and from the STAR-RIS to ${U_\chi}$, respectively, and ${{\vartheta _0}}$ and ${{\vartheta _\chi }}$ denote the corresponding path loss exponents.

In the transmission period, all the elements of the STAR-RIS are in transmission mode. The AP transmits the signals to ${U_t}$ in the downlink.
The received signal at ${U_t}$ is given by
\vspace{-0.3cm}
\begin{equation}
\label{eq:TSyt}
{y_{t,TEP}} \!=\! \sqrt {{P_{AP}}{{l_t}}} {\mathbf{\tilde {h}}}\mathrm{diag} ({\varphi ^{TEP}_{ET}}){\mathbf{\tilde {g}_t}}^T{s_{AP}} + {n_{t,TEP}}
 \!=\! \sqrt {{P_{AP}}{{l_t}}} \left( {\sum\limits_{i = 1}^N {{\tilde{h}_i}{e^{j{\varphi _i^{TEP}}}}{\tilde{g}_{t,i}}} } \right){s_{AP}} + {n_{t,TEP}},
 \vspace{-0.1cm}
\end{equation}
where ${{P_{AP}}}$ denotes the transmitted power at the AP, diag($\mathbf{X}$) is a diagonal matrix whose diagonals are the elements of $\mathbf{X}$, ${\mathbf{x}^T}$ denotes the transpose of vector $\mathbf{x}$, ${s_{AP}}$ is the signal that the AP transmits to ${U_t}$ with $\mathrm{E}\{ {\left| {{s_{AP}}} \right|^2}\}  \!=\! 1$, where $\mathrm{E}\left(  \cdot  \right)$ represents the expectation operation,
${n_{t,TEP}} \sim {\cal C}{\cal N}\left( {0,{N_0}} \right)$ is the additive white Gaussian noise (AWGN), where $\mathcal{CN}\left( {\mu ,{\sigma ^2}} \right)$ denotes the complex Gaussian distribution with mean $\mu$ and variance ${{\sigma ^2}}$.

The available energy at ${U_t}$ can be calculated as
\vspace{-0.3cm}
\begin{equation}
\label{eq:Xt}
{X_{t,TEP}} = {P_{AP}}{l_t}{\left| {\sum\limits_{i = 1}^N {{{\tilde h}_i}{e^{j\varphi _i^{TEP}}}{{\tilde g}_{t,i}}} } \right|^2}{\alpha _t}.
 \vspace{-0.3cm}
\end{equation}

Moreover, the \textit{complex channel coefficients} can be expressed in
polar coordinates as ${{\tilde h}_i} = {h_i}{e^{ - j{\mu _i}}}$ and ${{\tilde g}_{t,i}} = {g_i}{e^{ - j{\sigma _i}}}$, where ${h_i}$ and ${g_{t,i}}$ are \textit{the magnitudes
of the channel coefficients}, i.e., $\left| {{{\tilde h}_i}} \right| = {h_i}$ and $\left| {{{\tilde g}_{t,i}}} \right| = {g_{t,i}}$, and ${{\mu _i}}$ and ${{\sigma _i}}$ are are \textit{the phases} of ${{\tilde h}_i}$ and ${{\tilde g}_{t,i}}$, with $\left\{ {{\mu _i},{\sigma _i}} \right\} \in \left[ {0,2\pi } \right]$.
The magnitudes of the channel coefficients follow a Nakagami-$m$ distribution, i.e., ${h_i}\sim {\rm{Nakagami}}\left( {{m_i},{\Omega _i}} \right)$ and ${g_{t,i}}\sim {\rm{Nakagami}}\left( {{m_{t,i}},{\Omega _{t,i}}} \right)$, where ${m_i}$, ${m_{t,i}}$ are the shape parameters that are larger than $0$ and ${\Omega _i}$ and ${\Omega _{t,i}}$ are the spread parameters of the distribution.
Hence, (\ref{eq:Xt}) can be expressed as
\vspace{-0.3cm}
\begin{equation}
\label{eq:Xt1}
{X_{t,TEP}} = {P_{AP}}{l_t}{\left| {\sum\limits_{i = 1}^N {{h_i}{e^{j\left( {\varphi _i^{TEP} - {\mu _i} - {\sigma _i}} \right)}}{g_{t,i}}} } \right|^2}{\alpha _t}.
 \vspace{-0.3cm}
\end{equation}

To overcome the destructive effect of multipath fading, the phase-shifts of the STAR-RIS are reconfigured to obtain the maximum  available energy. Considering perfect channel state information (CSI)\cite{9740451,9570143}, the phase-shift of the STAR-RIS is set as $\varphi _i^{TEP} = {\mu _i} + {\sigma _i}$.
Thus, (\ref{eq:Xt1}) can be re-expressed as
${X_{t,TEP}} = {P_{AP}}{l_t}{\left| {\sum\limits_{i = 1}^N {{h_i}{g_{t,i}}} } \right|^2}{\alpha _t}.$

In the reflection period, all the elements of the STAR-RIS are in reflection mode. Similarly, the received signal at ${U_r}$ is given by
 \vspace{-0.5cm}
\begin{equation}
{y_{r,TEP}}{ =\! \sqrt {{P_{AP}}{{l_r}}} {\mathbf{\tilde{h}}}\mathrm{diag}({\zeta ^{TEP}_{ET}})\mathbf{\tilde{g}_r}^T{s_{AP}} + n_{r,TEP}}
 \!=\! \sqrt {{P_{AP}}{{l_r}}} \left( {\sum\limits_{i = 1}^N {{\tilde{h}_i}{e^{j\zeta _i^{TEP}}}{\tilde{g}_{r,i}}} }\! \right){s_{AP}} + n_{r,TEP},
 \vspace{-0.5cm}
\end{equation}
where ${n_{r,TEP}}$$\sim$$\mathcal{CN}\left( {0,{N_0}} \right)$.
The available energy is ${X_{r,TEP}} = {P_{AP}}{l_r}{\left| {\sum\limits_{i = 1}^N {{h_i}{g_{r,i}}} } \right|^2}{\alpha _r},$
where ${g_{r,i}}$ also follows ${\rm{Nakagami}}\left( {{m_{r,i}},{\Omega _{r,i}}} \right)$ with $m_{r,i}$ and $\Omega_{r,i}$ being the corresponding shape parameter and spread parameter of the distribution, respectively.

\subsubsection{IT Phase}
For the energy-splitting protocol, all the elements of the STAR-RIS work simultaneously in transmission and reflection modes with the energy-splitting ratios, i.e., $\sqrt{\beta_{t,TEP}}$ and $\sqrt {\beta_{r,TEP}}  \in \left[ {0,1} \right]$.
Accordingly, ${\boldsymbol{\varphi}^{TEP}_{IT}} \buildrel \Delta \over = \left[ {\sqrt{\beta_{t,TEP}}{e^{j{\varphi  _1^{TEP}}}}, \ldots ,\sqrt{\beta_{t,TEP}}{e^{j{\varphi _N^{TEP}}}}} \right]$ and ${\boldsymbol{\zeta}^{TEP}_{IT}} \buildrel \Delta \over = \left[ {\sqrt{\beta_{r,TEP}}{e^{j{\zeta _1^{TEP}}}}, \ldots ,\sqrt{\beta_{r,TEP}}{e^{j{\zeta _N^{TEP}}}}} \right]$ denote the transmission and reflection coefficient vectors of the STAR-RIS, respectively.
According to the law of energy conservation, $\beta_{t,TEP}+ \beta_{r,TEP} = 1$ holds\cite{9740451,9664576}.

In the IT phase, for the uplink NOMA system considered,
${U_t}$ and ${U_r}$ transmit data to the AP using the harvested energy over the same time and frequency resource block.
The received signal at the AP is given by
\vspace{-0.2cm}
\begin{equation}
\label{eq:yapP1}
{y_{AP,TEP}} \!=\! \sqrt {\frac{{{X_{t,TEP}}}}{{{\alpha _{AP}}}}{{l_t}}{{\beta _{t,TEP}}}}  \left( \!{\sum\limits_{i = 1}^N {{\tilde{h}_i}{e^{j\varphi _i^{TEP}}}{\tilde{g}_{t,i}}} } \!\right)\!{s_t}
 + \sqrt {\frac{{{X_{r,TEP}}}}{{{\alpha _{AP}}}}{{l_r}}{{\beta _{r,TEP}}}} \left(\! {\sum\limits_{i = 1}^N {{\tilde{h}_i}{e^{j\zeta _i^{TEP}}}{\tilde{g}_{r,i}}} } \!\right)\!{s_r} + {n_{AP}},
 \vspace{-0.5cm}
\end{equation}
where ${s_\chi }$ is the signal from ${U_\chi}$ with $\mathrm{E}\{ {\left| {{s_\chi }} \right|^2}\}  = 1$ and ${n_{AP}}$$\sim$$\mathcal{CN}\left( {0,{N_0}} \right)$.

The received signal-to-noise ratios (SNRs) of ${s_t}$ and ${s_r}$ are expressed as
\vspace{-0.2cm}
\begin{eqnarray}
{\gamma _{t,TEP}} = \frac{{{P_{AP}}l_t^2{\beta _{t,TEP}}{{\left| {\sum\limits_{i = 1}^N {{h_i}{g_{t,i}}} } \right|}^4}{\alpha _t}}}{{{\alpha _{AP}}{N_0}}},{\gamma _{r,TEP}} = \frac{{{P_{AP}}l_r^2{\beta _{r,TEP}}{{\left| {\sum\limits_{i = 1}^N {{h_i}{g_{r,i}}} } \right|}^4}{\alpha _r}}}{{{\alpha _{AP}}{N_0}}}.
\vspace{-0.7cm}
\end{eqnarray}
\vspace{-1.3cm}

The optimal decoding order policy for SIC is used to decode the received signal at the AP, given by\cite{9134947}
\vspace{-0.5cm}
\begin{spacing}{1}
\begin{align}
\label{eq:decode}
S = \left\{ {\begin{array}{*{20}{c}}
{\begin{array}{*{20}{c}}
{\left( {t,r} \right),}\\
{\left( {r,t} \right),}\\
{\left( {t,r} \right)\text{or}\left( {r,t} \right),}
\end{array}}&{\begin{array}{*{20}{c}}
{\frac{{{\gamma _{t,TEP}}}}{{{\gamma _{r,TEP}} + 1}} \ge {\gamma _{th}}\& \frac{{{\gamma _{r,TEP}}}}{{{\gamma _{t,TEP}} + 1}} < {\gamma _{th}}}\\
{\frac{{{\gamma _{t,TEP}}}}{{{\gamma _{r,TEP}} + 1}} < {\gamma _{th}}\& \frac{{{\gamma _{r,TEP}}}}{{{\gamma _{t,TEP}} + 1}} \ge {\gamma _{th}}}\\
{{\text{otherwise,}}}
\end{array}}
\end{array}} \right.
\vspace{-0.2cm}
\end{align}
\end{spacing}
where ${\gamma _{th}} = {2^R} - 1$ with $R$ being the target rate,
$S = \left( {x,y} \right)$ denotes the decoding order that $U_x$ is decoded before $U_y$.

 \vspace{-0.3cm}
\subsection{EEP}
For the EEP scheme, ${\boldsymbol{\varphi} ^{EEP}} \buildrel \Delta \over = \left[ {\sqrt {{\beta _{t,EEP}}} {e^{j\varphi _1^{EEP}}}, \ldots ,\sqrt {{\beta _{t,EEP}}} {e^{j\varphi _N^{EEP}}}} \right]$ and ${\boldsymbol{\zeta} ^{EEP}} \buildrel \Delta \over = \\ \left[ {\sqrt{\beta_{r,EEP}}{e^{j{\zeta _1^{EEP}}}},\ldots, \sqrt{\beta_{r,EEP}}{e^{j{\zeta _N^{EEP}}}}} \right]$ are defined as the transmission and reflection coefficient vectors of the STAR-RIS, respectively.
\subsubsection{ET phase}
The received signal at ${U_t}$ is given by
\vspace{-0.3cm}
\begin{align}
\label{eq:ESyt}
{y_{t,EEP}} = \sqrt {{P_{AP}}{{l_t}}{{\beta _{t,EEP}}}}  \left( {\sum\limits_{i = 1}^N {{\tilde{h}_i}{e^{j\varphi _i^{EEP}}}{\tilde{g}_{t,i}}} } \right){s_{AP}} + {n_{t,EEP}}.
 \vspace{-0.7cm}
\end{align}
The available energy at ${U_t}$ can be calculated as
\vspace{-0.6cm}
\begin{equation}
{X_{t,EEP}} = {P_{AP}}{l_t}{\beta _{t,EEP}}{\left| {\sum\limits_{i = 1}^N {{h_i}{g_{t,i}}} } \right|^2}{\alpha _{ET}}.
\vspace{-0.6cm}
\end{equation}

Similarly, at ${U_r}$, one has
\vspace{-0.3cm}
\begin{align}
\label{eq:ESyr}
{y_{r,EEP}}
 = \sqrt {{P_{AP}}{{l_r}}{{\beta _{r,EEP}}}}  \left( {\sum\limits_{i = 1}^N {{\tilde{h}_i}{e^{j\zeta _i^{EEP}}}{\tilde{g}_{r,i}}} } \right){s_{AP}} + {n_{r,EEP}},
 \vspace{-0.4cm}
\end{align}
and
\vspace{-0.3cm}
\begin{equation}
{X_{r,EEP}} = {P_{AP}}{l_r}{\beta _{r,EEP}}{\left| {\sum\limits_{i = 1}^N {{h_i}{g_{r,i}}} } \right|^2}{\alpha _{ET}}.
\vspace{-0.4cm}
\end{equation}

\subsubsection{IT Phase}
Similar to (\ref{eq:yapP1}), the received signal at the AP can be expressed as
\vspace{-0.1cm}
\begin{equation}
{y_{AP,EEP}} \!=\! \sqrt {\frac{{{X_{t,EEP}}}}{{{\alpha _{IT}}}}{{l_t}}{{\beta _{t,EEP}}}} \left(\! {\sum\limits_{i = 1}^N {{\tilde{h}_i}{e^{j\varphi _i^{EEP}}}{\tilde{g}_{t,i}}} } \!\right)\!{s_t}
 + \sqrt {\frac{{{X_{r,EEP}}}}{{{\alpha _{IT}}}}{{l_r}}{{\beta _{r,EEP}}}} \left(\! {\sum\limits_{i = 1}^N {{\tilde{h}_i}{e^{j\zeta _i^{EEP}}}{\tilde{g}_{r,i}}} } \!\right)\!{s_r} + {n_{AP}}.
 \vspace{-0.6cm}
\end{equation}

The received SNRs of ${s_t}$ and ${s_r}$ are written as
\vspace{-0.1cm}
\begin{eqnarray}
{\gamma _{t,EEP}} = \frac{{{P_{AP}}l_t^2\beta _{t,EEP}^2{{\left| {\sum\limits_{i = 1}^N {{h_i}{g_{t,i}}} } \right|}^4}{\alpha _{ET}}}}{{{\alpha _{IT}}{N_0}}},{\gamma _{r,EEP}} = \frac{{{P_{AP}}l_r^2\beta _{r,EEP}^2{{\left| {\sum\limits_{i = 1}^N {{h_i}{g_{r,i}}} } \right|}^4}{\alpha _{ET}}}}{{{\alpha _{IT}}{N_0}}}.
 \vspace{-0.8cm}
\end{eqnarray}

Similar to the TEP scheme, for the uplink NOMA system considered, the EEP also uses the optimal decoding order policy for SIC.
One has
\begin{spacing}{1}
\vspace{-0.2cm}
\begin{align}
\label{eq:Dr}
S = \left\{ {\begin{array}{*{20}{c}}
{\begin{array}{*{20}{c}}
{\left( {t,r} \right),}\\
{\left( {r,t} \right),}\\
{\left( {t,r} \right)\text{or}\left( {r,t} \right),}
\end{array}}&{\begin{array}{*{20}{c}}
{\frac{{{\gamma _{t,EEP}}}}{{{\gamma _{r,EEP}} + 1}} \ge {\gamma _{th}}\& \frac{{{\gamma _{r,EEP}}}}{{{\gamma _{t,EEP}} + 1}} < {\gamma _{th}}}\\
{\frac{{{\gamma _{t,EEP}}}}{{{\gamma _{r,EEP}} + 1}} < {\gamma _{th}}\& \frac{{{\gamma _{r,EEP}}}}{{{\gamma _{t,EEP}} + 1}} \ge {\gamma _{th}}}\\
{{\text{otherwise.}}}
\end{array}}
\end{array}} \right.
 \vspace{-0.6cm}
\end{align}
\end{spacing}
 \vspace{-0.3cm}
\subsection{STAR-RIS Aided Wireless-Powered TDMA System}
For OMA schemes, users are served by TDMA to avoid inter-user interference. The STAR-RIS operates in the time-switching protocol
to support the downlink and uplink transmissions in a TDMA way. The ET phase is divided into ${\alpha _{t}}{T_s}$ for $U_t$ and ${\alpha _{r}}{T_s}$ for $U_r$ in the downlink, while the IT phase is divided into $\alpha _{AP}^t{T_s}$ for $U_t$ and $\alpha _{AP}^r{T_s}$ for $U_r$ in the uplink, where $\alpha_{AP}^t +\alpha_{AP}^r = 1 - \alpha_t - \alpha_r$. Hence, for the STAR-RIS aided wireless-powered TDMA system, the received SNR at ${U_\chi}$ is computed as
\begin{eqnarray}
 \vspace{-1.6cm}
{\gamma _\chi } = \frac{{{P_{AP}}l_\chi ^2{{\left| {\sum\limits_{i = 1}^N {{h_i}{g_{\chi ,i}}} } \right|}^4}{\alpha _\chi }}}{{\alpha _{AP}^\chi {N_0}}}.
 \vspace{-0.1cm}
\end{eqnarray}

\section{Outage Probability and Sum Throughput} \label{sect:perfomnanceanalysis}
In this section, the outage probability and sum throughput of the proposed system with the TEP and EEP, and the STAR-RIS aided wireless-powered TDMA system are derived over Nakagami-$m$ fading channels.

Let ${H_{{h_i}{g_{t,i}}}} = {h_i}{g_{t,i}}$, the $n$-th moment of ${H_{{h_i}{g_{t,i}}}}$ is expressed as
\vspace{-0.4cm}
\begin{equation}
\label{eq:kth}
{\mu _{{H_{{h_i}{g_{t,i}}}}}}\left( n \right) = \lambda _{{h_i}{g_{t,i}}}^{ - n}\frac{{\Gamma \left( {{m_i} + n/2} \right)\Gamma \left( {{m_{t,i}} + n/2} \right)}}{{\Gamma \left( {{m_i}} \right)\Gamma \left( {{m_{t,i}}} \right)}},
\vspace{-0.2cm}
\end{equation}
where ${\lambda _{{h_i}{g_{t,i}}}} = \sqrt {\frac{{{m_i}}}{{{\Omega _i}}}\frac{{{m_{t,i}}}}{{{\Omega _{t,i}}}}}$, $\Gamma \left(  \cdot  \right)$ is the Gamma function.
The detailed dervation of (\ref{eq:kth}) is provided in Appendix A.

Here, the moment-matching approach is used to approximate the distribution of ${H_{{h_i}{g_{t,i}}}}$ with a Gamma distribution\cite{9138463,9774334}, i.e.,
${H_{{h_i}{g_{t,i}}}} \sim Gamma\left( {k,\theta } \right),$
where the $k$ and $\theta$ is expressed as
\vspace{-0.3cm}
\begin{eqnarray}
k \!=\! \frac{{{\mathrm{E}^2}\left( {{H_{{h_i}{g_{t,i}}}}} \right)}}{{\mathrm{Var}\left( {{H_{{h_i}{g_{t,i}}}}} \right)}} = \frac{{\mu _{{H_{{h_i}{g_{t,i}}}}}^2\left( 1 \right)}}{{{\mu _{{H_{{h_i}{g_{t,i}}}}}}\left( 2 \right) - \mu _{{H_{{h_i}{g_{t,i}}}}}^2\left( 1 \right)}},\theta  \!=\! \frac{{ \mathrm{E}\left( {{H_{{h_i}{g_{t,i}}}}} \right)}}{{\mathrm{Var}\left( {{H_{{h_i}{g_{t,i}}}}} \right)}} = \frac{{{\mu _{{H_{{h_i}{g_{t,i}}}}}}\left( 1 \right)}}{{{\mu _{{H_{{h_i}{g_{t,i}}}}}}\left( 2 \right) - \mu _{{H_{{h_i}{g_{t,i}}}}}^2\left( 1 \right)}},
\vspace{-0.1cm}
\end{eqnarray}
$\mathrm{Var}\left(  \cdot  \right)$ represents the variance operation.

Let ${G_{{h_i}{g_{t,i}}}} = \sum\limits_{i = 1}^N {{h_i}{g_{t,i}}}$, one has
${G_{{h_i}{g_{t,i}}}} \sim Gamma\left( {Nk,\theta } \right).$
The cumulative distribution function (CDF) of ${G_{{h_i}{g_{t,i}}}}$ is expressed as
\vspace{-0.4cm}
\begin{equation}
{F_{{G_{{h_i}{g_{t,i}}}}}}\left( x \right) = \frac{1}{{\Gamma \left( {Nk} \right)}}\gamma \left( {Nk,\theta x} \right),
\vspace{-0.2cm}
\end{equation}
where $\gamma \left( { \cdot , \cdot } \right)$ is the upper incomplete Gamma function.

For $Y = {X^2}$, the CDF of $Y$ can be calculated as ${F_Y}\left( y \right) = {F_X}\left( {\sqrt {{y}} } \right)$. The CDF of ${\left| {{G_{{h_i}{g_{t,i}}}}} \right|^4}$ is computed as
\vspace{-0.3cm}
\begin{equation}
\label{eq:CDFtP1}
{F_{{{\left| {{G_{{h_i}{g_{t,i}}}}} \right|}^4}}}\left( x \right) = \frac{1}{{\Gamma \left( {Nk} \right)}}\gamma \left( {Nk,\theta {x^{\frac{1}{4}}}} \right)
= 1 - \left( {{e^{ - \theta {x^{\frac{1}{4}}}}} + \sum\limits_{m = 1}^{Nk - 1} {{e^{ - \theta {x^{\frac{1}{4}}}}}\frac{{{\theta ^m}{x^{\frac{m}{4}}}}}{{m!}}} } \right).
\vspace{-0.1cm}
\end{equation}

The probability density function (PDF) of ${\left| {{G_{{h_i}{g_{t,i}}}}} \right|^4}$ is obtained as
\vspace{-0.3cm}
\begin{equation}
\label{eq:fxtP1}
{f_{{{\left| {{G_{{h_i}{g_{t,i}}}}} \right|}^4}}}\left( x \right) = \frac{{{\theta ^{Nk}}{e^{ - \theta {x^{\frac{1}{4}}}}}{x^{\frac{{Nk - 4}}{4}}}}}{{4\left( {Nk - 1} \right)!}}.
\vspace{-0.1cm}
\end{equation}

Let ${G_{{h_i}{g_{r,i}}}} = \sum\limits_{i = 1}^N {{h_i}{g_{r,i}}}$. The PDF and CDF of ${G_{{h_i}{g_{r,i}}}}$ are the same as those of ${G_{{h_i}{g_{t,i}}}}$. Hence, one has
\begin{spacing}{1}
\vspace{-0.3cm}
\begin{equation}
F(x) = {F_{{{\left| {{G_{{h_i}{g_{r,i}}}}} \right|}^4}}}\left( x \right) = {F_{{{\left| {{G_{{h_i}{g_{t,i}}}}} \right|}^4}}}\left( x \right)
 = \frac{1}{{\Gamma \left( {Nk} \right)}}\gamma \left( {Nk,\theta {x^{\frac{1}{4}}}} \right),
\vspace{-0.1cm}
\end{equation}
\begin{equation}
f(x) = {f_{{{\left| {{G_{{h_i}{g_{r,i}}}}} \right|}^4}}}\left( x \right) = {f_{{{\left| {{G_{{h_i}{g_{t,i}}}}} \right|}^4}}}\left( x \right)
 = \frac{{{\theta ^{Nk}}{e^{ - \theta {x^{\frac{1}{4}}}}}{x^{\frac{{Nk - 4}}{4}}}}}{{4\left( {Nk - 1} \right)!}}.
\vspace{-0.1cm}
\end{equation}
\end{spacing}
 \vspace{-0.3cm}
\subsection{TEP}
According to the optimal decoding order policy in (\ref{eq:decode}), the outage probability
of $U_t$ for the proposed system with the TEP scheme is given by
\vspace{-0.4cm}
\begin{small}
\begin{align}
\label{eq:Poutt}
P_{out,t}^{{\rm{TEP}}} &= \Pr \left( {\frac{{{\gamma _{t,{\rm{TEP}}}}}}{{{\gamma _{r,{\rm{TEP}}}} + 1}} < {\gamma _{th}},\frac{{{\gamma _{r,{\rm{TEP}}}}}}{{{\gamma _{t,{\rm{TEP}}}} + 1}} < {\gamma _{th}}} \right) + \Pr \left( {{\gamma _{t,{\rm{TEP}}}} < {\gamma _{th}},\frac{{{\gamma _{r,{\rm{TEP}}}}}}{{{\gamma _{t,{\rm{TEP}}}} + 1}} \ge {\gamma _{th}}} \right)\nonumber\\
& = F\left( {\frac{{{\gamma _{th}}}}{B}} \right) + \sum\limits_{m = 0}^{Nk - 1} {\frac{{{\theta ^{Nk + m}}}}{{4m!\left( {Nk - 1} \right)!}}} \nonumber\\
& \times \!\!\left( \!\!\!\begin{array}{l}
\sum\limits_{w = 1}^W {{\psi _w}} {e^{ - \theta {{\left( {\frac{{B{e^{{u_w}}} + {\gamma _{th}}}}{{A{\gamma _{th}}}} - \frac{1}{A}} \right)}^{\frac{1}{4}}}}}{\left( {\frac{{B{e^{{u_w}}} + {\gamma _{th}}}}{{A{\gamma _{th}}}} - \frac{1}{A}} \right)^{\frac{m}{4}}}{e^{ - \theta {{\left( {{e^{{u_w}}} + \frac{{{\gamma _{th}}}}{B}} \right)}^{\frac{1}{4}}}}}{\left( {{e^{{u_w}}} + \frac{{{\gamma _{th}}}}{B}} \right)^{\frac{{Nk - 4}}{4}}}{e^{{u_w} + u_w^2}}\\
 - \sum\limits_{w = 1}^W {{\psi _w}} {e^{ - \theta {{\left( {\frac{{{\gamma _{th}}}}{A}\left( {B{e^{{u_w}}} + 1} \right)} \right)}^{\frac{1}{4}}}}}{\left( {\frac{{{\gamma _{th}}}}{A}\left( {B{e^{{u_w}}} + 1} \right)} \right)^{\frac{m}{4}}}{e^{ - \theta {e^{\frac{{{u_w}}}{4}}}}}{e^{\frac{{{u_w}Nk}}{4} + u_w^2}}\\
 + \int\limits_0^{\frac{{{\gamma _{th}}}}{A}}  {e^{ - \theta {{\left( {\frac{{{\gamma _{th}}}}{B}\left( {Ax + 1} \right)} \right)}^{\frac{1}{4}}}}}{\left( {\frac{{{\gamma _{th}}}}{B}\left( {Ax + 1} \right)} \right)^{\frac{m}{4}}}{e^{ - \theta {x^{\frac{1}{4}}}}}{x^{\frac{{Nk - 4}}{4}}}dx
\end{array} \!\!\!\right)\!\!,
\end{align}
\end{small}
where $A = \frac{{{P_{AP}}l_t^2{\beta _{t,TEP}}{\alpha _t}}}{{{\alpha _{AP}}{N_0}}}$, $B = \frac{{{P_{AP}}l_r^2{\beta _{r,TEP}}{\alpha _r}}}{{{\alpha _{AP}}{N_0}}}$. The detailed derivation of (\ref{eq:Poutt}) is provided in Appendix B.

Similarly, the outage probability of ${U_r}$ for the proposed system with the TEP scheme can be written as
\vspace{-0.4cm}
\begin{spacing}{1}
\begin{small}
\begin{align}
\label{eq:Poutr}
P_{out,r}^{{\rm{TEP}}} &= \Pr \left( {\frac{{{\gamma _{r,{\rm{TEP}}}}}}{{{\gamma _{t,{\rm{TEP}}}} + 1}} < {\gamma _{th}},\frac{{{\gamma _{t,{\rm{TEP}}}}}}{{{\gamma _{r,{\rm{TEP}}}} + 1}} < {\gamma _{th}}} \right) + \Pr \left( {{\gamma _{r,{\rm{TEP}}}} < {\gamma _{th}},\frac{{{\gamma _{t,{\rm{TEP}}}}}}{{{\gamma _{r,{\rm{TEP}}}} + 1}} \ge {\gamma _{th}}} \right)\nonumber \\
 &= F\left( {\frac{{{\gamma _{th}}}}{A}} \right) + \sum\limits_{m = 0}^{Nk - 1} {\frac{{{\theta ^{Nk + m}}}}{{4m!\left( {Nk - 1} \right)!}}}\nonumber  \\
 &\times \!\!\left( \!\!\!\begin{array}{l}
\sum\limits_{w = 1}^W {{\psi _w}} {e^{ - \theta {{\left( {\frac{{A{e^{{u_w}}} + {\gamma _{th}}}}{{B{\gamma _{th}}}} - \frac{1}{B}} \right)}^{\frac{1}{4}}}}}{\left( {\frac{{A{e^{{u_w}}} + {\gamma _{th}}}}{{B{\gamma _{th}}}} - \frac{1}{B}} \right)^{\frac{m}{4}}}{e^{ - \theta {{\left( {{e^{{u_w}}} + \frac{{{\gamma _{th}}}}{A}} \right)}^{\frac{1}{4}}}}}{\left( {{e^{{u_w}}} + \frac{{{\gamma _{th}}}}{A}} \right)^{\frac{{Nk - 4}}{4}}}{e^{{u_w} + u_w^2}}\\
 - \sum\limits_{w = 1}^W {{\psi _w}} {e^{ - \theta {{\left( {\frac{{{\gamma _{th}}}}{B}\left( {A{e^{{u_w}}} + 1} \right)} \right)}^{\frac{1}{4}}}}}{\left( {\frac{{{\gamma _{th}}}}{B}\left( {A{e^{{u_w}}} + 1} \right)} \right)^{\frac{m}{4}}}{e^{ - \theta {e^{\frac{{{u_w}}}{4}}}}}{e^{\frac{{{u_w}Nk}}{4} + u_w^2}}\\
 + \int\limits_0^{\frac{{{\gamma _{th}}}}{B}} {{e^{ - \theta {{\left( {\frac{{{\gamma _{th}}}}{A}\left( {Bx + 1} \right)} \right)}^{\frac{1}{4}}}}}} {\left( {\frac{{{\gamma _{th}}}}{A}\left( {Bx + 1} \right)} \right)^{\frac{m}{4}}}{e^{ - \theta {x^{\frac{1}{4}}}}}{x^{\frac{{Nk - 4}}{4}}}dx
\end{array} \!\!\!\right)\!\!.
\vspace{-0.2cm}
\end{align}
\end{small}
\end{spacing}
Finally, the sum throughput of the proposed system with the TEP scheme can be expressed as
\vspace{-0.4cm}
\begin{equation}
\label{eq:sumthP1}
T^{{TEP}} = R{\alpha _{{\rm{AP}}}}\left( {1 - P_{out,t}^{{TEP}}} \right) + R{\alpha _{{\rm{AP}}}}\left( {1 - P_{out,r}^{{TEP}}} \right).
\vspace{-0.2cm}
\end{equation}

 \vspace{-0.8cm}

\subsection{EEP}
According to (\ref{eq:Dr}), the outage probability of the proposed system with the EEP scheme is calculated as
\vspace{-0.4cm}
\begin{small}
\begin{align}
P_{out,\chi }^{{\rm{EEP}}} &= F\left( {\frac{{{\gamma _{th}}}}{{{D_\chi }}}} \right) + \sum\limits_{m = 0}^{Nk - 1} {\frac{{{\theta ^{Nk + m}}}}{{4m!\left( {Nk - 1} \right)!}}} \nonumber\\
 &\times\!\! \left( \!\!\!\begin{array}{l}
\sum\limits_{w = 1}^W {{\psi _w}} {e^{ - \theta {{\left( {\frac{{{D_\chi }{e^{{u_w}}} + {\gamma _{th}}}}{{{C_\chi }{\gamma _{th}}}} - \frac{1}{{{C_\chi }}}} \right)}^{\frac{1}{4}}}}}\!\!{\left(\!\! {\frac{{{D_\chi }{e^{{u_w}}} + {\gamma _{th}}}}{{{C_\chi }{\gamma _{th}}}} - \frac{1}{{{C_\chi }}}} \!\!\right)^{\frac{m}{4}}}\!\!{e^{ - \theta {{\left( {{e^{{u_w}}} + \frac{{{\gamma _{th}}}}{{{D_\chi }}}} \right)}^{\frac{1}{4}}}}}\!\!{\left( \! {{e^{{u_w}}} + \frac{{{\gamma _{th}}}}{{{D_\chi }}}} \!\right)^{\frac{{Nk - 4}}{4}}}\!\!{e^{{u_w} + u_w^2}}\\
 - \sum\limits_{w = 1}^W {{\psi _w}} {e^{ - \theta {{\left( {\frac{{{\gamma _{th}}}}{{{C_\chi }}}\left( {{D_\chi }{e^{{u_w}}} + 1} \right)} \right)}^{\frac{1}{4}}}}}{\left( {\frac{{{\gamma _{th}}}}{{{C_\chi }}}\left( {{D_\chi }{e^{{u_w}}} + 1} \right)} \right)^{\frac{m}{4}}}{e^{ - \theta {e^{\frac{{{u_w}}}{4}}}}}{e^{\frac{{{u_w}Nk}}{4} + u_w^2}}\\
 + \int\limits_0^{\frac{{{\gamma _{th}}}}{{{C_\chi }}}} {{e^{ - \theta {{\left( {\frac{{{\gamma _{th}}}}{{{D_\chi }}}\left( {{C_\chi }x + 1} \right)} \right)}^{\frac{1}{4}}}}}} {\left( {\frac{{{\gamma _{th}}}}{{{D_\chi }}}\left( {{C_\chi }x + 1} \right)} \right)^{\frac{m}{4}}}{e^{ - \theta {x^{\frac{1}{4}}}}}{x^{\frac{{Nk - 4}}{4}}}dx
\end{array} \!\!\!\right)\!\!,
\end{align}
\end{small}
where ${C_t} = \frac{{{P_{AP}}l_t^2\beta _{t,EEP}^2{\alpha _{ET}}}}{{{\alpha _{IT}}{N_0}}}$ and ${D_t} = \frac{{{P_{AP}}l_r^2\beta _{r,EEP}^2{\alpha _{ET}}}}{{{\alpha _{IT}}{N_0}}}$ for $U_t$, and ${C_r} = \frac{{{P_{AP}}l_r^2\beta _{r,EEP}^2{\alpha _{ET}}}}{{{\alpha _{IT}}{N_0}}}$ and ${D_r} = \frac{{{P_{AP}}l_t^2\beta _{t,EEP}^2{\alpha _{ET}}}}{{{\alpha _{IT}}{N_0}}}$ for $U_r$.

Finally, the sum throughput of the proposed system with the EEP scheme can be expressed as
\vspace{-0.4cm}
\begin{equation}
\label{eq:sumthP2}
T^{{EEP}} = R{\alpha _{IT}}\left( {1 - P_{out,t}^{{EEP}}} \right) + R{\alpha _{IT}}\left( {1 - P_{out,r}^{{EEP}}} \right).
\vspace{-0.1cm}
\end{equation}
\subsection{STAR-RIS Aided Wireless-Powered TDMA System}
For the STAR-RIS aided wireless-powered TDMA system, the outage probability of ${U_\chi}$ is given by
\begin{spacing}{1}
\begin{small}
\vspace{-0.5cm}
\begin{align}
P_{out,\chi }^{{TDMA}}  &= \Pr \left( {{\gamma _\chi } < {\gamma _{th}}} \right)
 = \Pr \left( {\frac{{{P_{AP}}l_\chi ^2{{\left| {\sum\limits_{i = 1}^N {{h_i}{g_{\chi ,i}}} } \right|}^4}{\alpha _\chi }}}{{\alpha _{AP}^\chi {N_0}}} < {\gamma _{th}}} \right)\nonumber\\
 &= \Pr \left( {{{\left| {\sum\limits_{i = 1}^N {{h_i}{g_{\chi ,i}}} } \right|}^4} < \frac{{{\gamma _{th}}\alpha _{AP}^\chi {N_0}}}{{{P_{AP}}l_\chi ^2{\alpha _\chi }}}} \right)
 = F\left( {\frac{{{\gamma _{th}}\alpha _{AP}^\chi {N_0}}}{{{P_{AP}}l_\chi ^2{\alpha _\chi }}}} \right).
\end{align}
\end{small}
\end{spacing}

The sum throughput of the STAR-RIS aided wireless-powered TDMA system can be expressed as
\vspace{-0.8cm}
\begin{eqnarray}
T^{{TDMA}} = R\alpha _{AP}^t\left( 1 - P_{out,t }^{TDMA} \right) + R\alpha _{AP}^r\left( 1 - P_{out,r }^{TDMA} \right).
\vspace{-0.1cm}
\end{eqnarray}

\section{Age of Information}
In this section, the average AoIs of the proposed system and the STAR-RIS aided wireless-powered TDMA system are analyzed.
The duration of each communication process ${T_s}$, including the ET and IT phases, is employed to calculate the AoI.
In time slot $\tau $, the AoI of the system is defined as
\vspace{-0.3cm}
\begin{equation}
\Delta \left( \tau  \right) = \tau - U\left( \tau \right),
\vspace{-0.3cm}
\end{equation}
where $U\left( \tau \right)$ is the generation time of the most recently
received packet at AP.
\begin{figure}
\center
\includegraphics[width=3.2in,height=2in]{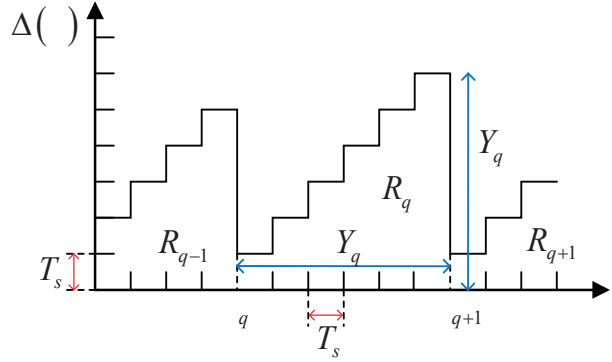}
\vspace{-0.5cm}
\captionsetup{font={small}}
\caption{Example of AoI.}
\label{fig:AoImodel}
\vspace{-1.1cm}
\end{figure}
Fig.~\ref{fig:AoImodel} shows an example of the age evolution for the proposed system, where
${\tau _q}$ represents the time slot of the last time successful update at AP, ${\tau _{q + 1}}$
represents the current time slot successfully updated at AP, and ${Y_q} = {\tau _{q + 1}} - {\tau _q}$  denotes the required time for the $q$-th successful update (time difference between ${\tau _{q + 1}}$ and ${\tau _q}$).
If the decoding is successful, ${\Delta \left( \tau  \right)}$ at AP is reset to one.

Firstly, we derive the first-order and second-order moments for the required time between two consecutive successful updates at AP, i.e., ${Y_q}$.
${Y_q} = M{{T_s}} $, where $M$ is a discrete random variable that denotes the number of consecutive communications until successful decoding. If $M$ transmissions occur, this means that the previous $(M-1)$ consecutive transmissions are unsuccessful, while the $M$-th transmission is successful\cite{8606155}. According to (\!\!\cite{Gradshte1965Table}, eq.~(1.113)), the expectation of ${Y_q}$ can be calculated as
\vspace{-0.3cm}
\begin{equation}
\label{eq:EY}
\mathrm{E}\left( Y_q \right) = \sum\limits_{M = 1}^\infty  {M\mathrm{E}\left( {{T_s}} \right)} {\left( {1 - \Phi   } \right)^{M - 1}}\Phi   = \frac{1}{\Phi  },
\vspace{-0.1cm}
\end{equation}
where $\mathrm{E}\left( {{T_s}} \right) = 1$ and $\Phi $ is the success probability.

For the second-order moment of $Y_q$, one has $Y_q^2 = {M^2}T_s^2$.
Since $M$ is a discrete random variable,
by taking the conditional expectation operator of ${Y_q^2}$, one obtains
${\rm{E}}\left( {\left. {Y_q^2} \right|M} \right) = {M^2}{\rm{E}}\left( {T_s^2} \right)$.

Similar to (\!\!\cite{8606155}, eq.~(20)), the second-order moment of $Y_q$ is calculated as
\vspace{-0.4cm}
\begin{eqnarray}
\label{eq:EY2}
{\rm{E}}\left( {Y_q^2} \right) = \sum\limits_{M = 1}^\infty  {{\rm{E}}\left( {\left. {Y_q^2} \right|M} \right)} {\left( {1 - \Phi } \right)^{M - 1}}\Phi  = {\rm{E}}\left( {T_s^2} \right)\left( {\frac{2}{{{\Phi ^2}}} - \frac{1}{\Phi }} \right) = \frac{2}{{{\Phi ^2}}} - \frac{1}{\Phi },
 \vspace{-0.2cm}
\end{eqnarray}

where $\mathrm{E}\left( {T_s^2} \right) = 1$.

For a time period of $\eta $ time slots in which $\kappa $ status updates occur, the average time ${\Delta _\eta}$ of the proposed system is given by
\vspace{-0.3cm}
\begin{equation}
{\Delta _\eta} = \frac{1}{\eta}\sum\limits_{\tau  = 1}^\eta {\Delta \left( \tau  \right) = } \frac{1}{\eta}\sum\limits_{q = 1}^\kappa  {{R_q} = \frac{\kappa }{\eta}} \frac{1}{\kappa}\sum\limits_{q = 1}^\kappa {{R_q}},
\vspace{-0.3cm}
\end{equation}
where ${{R_q}}$ denotes the area under ${\Delta \left( \tau  \right)}$ corresponding to the $q$-th status update. When $\eta$ tends to infinity, the average time ${\Delta _\eta}$ tends to the average AoI \cite{8187436}, i.e.,
\vspace{-0.3cm}
\begin{align}
\Delta  &= \mathop {\lim }\limits_{\eta  \to \infty } {\Delta _\eta } = \mathop {\lim }\limits_{\eta  \to \infty } \frac{\kappa }{\eta }\frac{1}{\kappa }\sum\limits_{q = 1}^\kappa  {{R_q}}
 = \mathop {\lim }\limits_{\eta  \to \infty } \frac{\kappa }{\eta }\frac{1}{\kappa }\kappa \mathrm{E}\left( {{R_q}} \right) = \frac{{\mathrm{E}\left( {{R_q}} \right)}}{{\mathrm{E}\left( {{Y_q}} \right)}},
 \vspace{-0.2cm}
\end{align}
where $\mathop {\lim }\limits_{\eta  \to \infty } \frac{\kappa }{\eta } = \frac{1}{{\mathrm{E}\left( {{Y_q}} \right)}}$\cite{8606155}.

The area ${{R_q}}$ consists of ${Y_q}$ rectangles. One side of the rectangle is one and another side is $\ell$, where $1 \le \ell  \le {Y_q}$. As shown in Fig.~\ref{fig:AoImodel}, one has ${R_q} = \sum\limits_{\ell  = 1}^{{Y_q}} {\ell  = \frac{{{Y_q}\left( {{Y_q} + 1} \right)}}{2}} .$
Taking the expectation of $R_q$, one has $\mathrm{E}\left( {{R_q}} \right) = \frac{{\mathrm{E}\left( {Y_q^2} \right) + \mathrm{E}\left( {{Y_q}} \right)}}{2}.$
Thus, the average AoI is calculated as
\vspace{-0.3cm}
\begin{align}
\label{eq:AoI}
 \Delta  = \frac{1}{2}\left( {\frac{{{\rm{E}}\left( {Y_q^2} \right)}}{{{\rm{E}}\left( {{Y_q}} \right)}} + 1} \right) = \frac{1}{2}\left( {\frac{{\frac{2}{{{\Phi ^2}}} - \frac{1}{\Phi }}}{{\frac{1}{\Phi }}} + 1} \right) = \frac{1}{\Phi }.
 \vspace{-0.3cm}
\end{align}

The success probability ${\Phi  _{TEP}}$ for the TEP scheme is given by
\vspace{-0.3cm}
\begin{small}
\begin{align}
\label{eq:TEP}
&{\Phi _{TEP}} = \sum\limits_{m = 0}^{Nk - 1} {\frac{{{\theta ^{Nk}}}}{{4\left( {Nk - 1} \right)!m!}}}\nonumber \\
& \times \left(\!\! \begin{array}{l}
\sum\limits_{w = 1}^W {{\psi _w}} {e^{ - {{\left( {\frac{{{\theta ^4}{\gamma _{th}}{U_3}}}{{{U_1}}}} \right)}^{\frac{1}{4}}}}}{\left( {\frac{{{\theta ^4}{\gamma _{th}}{U_3}}}{{{U_1}}}} \right)^{\frac{m}{4}}}{e^{ - \theta {{\left( {{e^{{u_w}}} + \frac{{{\gamma _{th}}{\alpha _{AP}}{N_0}}}{{{U_2}}}} \right)}^{\frac{1}{4}}}}}{\left( {{e^{{u_w}}} + \frac{{{\gamma _{th}}{\alpha _{AP}}{N_0}}}{{{U_2}}}} \right)^{\frac{{Nk - 4}}{4}}}{e^{{u_w} + u_w^2}}\\
 + \sum\limits_{w = 1}^W {{\psi _w}} {e^{ - {{\left( {\frac{{{\theta ^4}{\gamma _{th}}{U_4}}}{{{U_2}}}} \right)}^{\frac{1}{4}}}}}{\left( {\frac{{{\theta ^4}{\gamma _{th}}{U_4}}}{{{U_2}}}} \right)^{\frac{m}{4}}}{e^{ - \theta {{\left( {{e^{{u_w}}} + \frac{{{\gamma _{th}}{\alpha _{AP}}{N_0}}}{{{U_1}}}} \right)}^{\frac{1}{4}}}}}{\left( {{e^{{u_w}}} + \frac{{{\gamma _{th}}{\alpha _{AP}}{N_0}}}{{{U_1}}}} \right)^{\frac{{Nk - 4}}{4}}}{e^{{u_w} + u_w^2}}
\end{array}\!\! \right) \!\!+\! 2{O_W},
\end{align}
\end{small}
where ${U_1} = {P_{AP}}l_r^2{\beta _{r,TEP}}{\alpha _r}$, ${U_2} = {P_{AP}}l_t^2{\beta _{t,TEP}}{\alpha _t}$, ${U_3} = {U_2}{e^{{u_w}}} + {\gamma _{th}}{\alpha _{AP}}{N_0} + {\alpha _{AP}}{N_0}$, and ${U_4} = {U_1}{e^{{u_w}}} + {\gamma _{th}}{\alpha _{AP}}{N_0} + {\alpha _{AP}}{N_0}$. The detailed derivation of (\ref{eq:TEP}) is provided in Appendix C.

Similarly, the success probability $\Phi _{EEP}$ for the EEP scheme can be calculated as
\vspace{-0.2cm}
\begin{small}
\begin{align}
&{\Phi _{EEP}} = \sum\limits_{m = 0}^{Nk - 1} {\frac{{{\theta ^{Nk}}}}{{4\left( {Nk - 1} \right)!m!}}} \nonumber\\
 &\times\!\! \left(\!\! \begin{array}{l}
\sum\limits_{w = 1}^W {{\psi _w}} {e^{ - {{\left( {\frac{{{\theta ^4}{\gamma _{th}}({V_3})}}{{{V_1}}}} \right)}^{\frac{1}{4}}}}}{\left( {\frac{{{\theta ^4}{\gamma _{th}}({V_3})}}{{{V_1}}}} \right)^{\frac{m}{4}}}{e^{ - \theta {{\left( {{e^{{u_w}}} + \frac{{{\gamma _{th}}{\alpha _{IT}}{N_0}}}{{{V_2}}}} \right)}^{\frac{1}{4}}}}}{\left( {{e^{{u_w}}} + \frac{{{\gamma _{th}}{\alpha _{IT}}{N_0}}}{{{V_2}}}} \right)^{\frac{{Nk - 4}}{4}}}{e^{{u_w} + u_w^2}}\\
 + \sum\limits_{w = 1}^W {{\psi _w}} {e^{ - {{\left( {\frac{{{\theta ^4}{\gamma _{th}}({V_4})}}{{{V_2}}}} \right)}^{\frac{1}{4}}}}}{\left( {\frac{{{\theta ^4}{\gamma _{th}}({V_4})}}{{{V_2}}}} \right)^{\frac{m}{4}}}{e^{ - \theta {{\left( {{e^{{u_w}}} + \frac{{{\gamma _{th}}{\alpha _{IT}}{N_0}}}{{{V_1}}}} \right)}^{\frac{1}{4}}}}}{\left( {{e^{{u_w}}} + \frac{{{\gamma _{th}}{\alpha _{IT}}{N_0}}}{{{V_1}}}} \right)^{\frac{{Nk - 4}}{4}}}{e^{{u_w} + u_w^2}}
\end{array} \!\!\right) \!\!+\! 2{O_W},
\end{align}
\end{small}
where ${V_1} = {P_{AP}}l_r^2\beta _{r,EEP}^2{\alpha _{ET}}$, ${V_2} = {P_{AP}}l_t^2\beta _{t,EEP}^2{\alpha _{ET}}$, ${V_3} = {V_2}{e^{{u_w}}} + {\gamma _{th}}{\alpha _{IT}}{N_0} + {\alpha _{IT}}{N_0}$, and ${V_4} = {V_1}{e^{{u_w}}} + {\gamma _{th}}{\alpha _{IT}}{N_0} + {\alpha _{IT}}{N_0}$.

The success probability ${\Phi  _{T}}$ for the STAR-RIS aided wireless-powered TDMA system is given by
\vspace{-0.3cm}
\begin{small}
\begin{align}
\label{eq:TDMA}
{\Phi _T} &= 2 - \sum\limits_{m = 0}^{Nk - 1} {\frac{{{\theta ^{m + Nk}}{U_5}^{\frac{m}{4}}}}{{m!\left( {Nk - 1} \right)!}}} {\left( {\theta \left( {{U_5}^{\frac{1}{4}} + 1} \right)} \right)^{ - \left( {Nk + m} \right)}}\Gamma \left( {Nk + m} \right) - \sum\limits_{m = 0}^{Nk - 1} {\frac{{{\theta ^{m + Nk}}{U_6}^{\frac{m}{4}}}}{{m!\left( {Nk - 1} \right)!}}} \nonumber\\
& \times {\left( {\theta \left( {{U_6}^{\frac{1}{4}} + 1} \right)} \right)^{ - \left( {Nk + m} \right)}}\Gamma \left( {Nk + m} \right) - F\left( {\frac{{{\gamma _{th}}\alpha _{AP}^t{N_0}}}{{{P_{AP}}l_t^2{\alpha _t}}}} \right) - F\left( {\frac{{{\gamma _{th}}\alpha _{AP}^r{N_0}}}{{{P_{AP}}l_r^2{\alpha _r}}}} \right),
\vspace{-0.1cm}
\end{align}
\end{small}
where ${U_5} = \frac{{l_r^2{\alpha _r}\alpha _{AP}^t}}{{\alpha _{AP}^rl_t^2{\alpha _t}}}$, ${U_6} = \frac{{l_t^2{\alpha _t}\alpha _{AP}^r}}{{\alpha _{AP}^tl_r^2{\alpha _r}}}$. The detailed derivation of (\ref{eq:TDMA}) is provided in Appendix C.

\section{Resource Allocation}
In this section, resource allocation is analyzed to maximize the sum throughput within an average AoI constraint.

\subsection{Problem Formulation}

\subsubsection{TEP}
We formulate the sum throughput problem as an optimization problem seeking to jointly optimize the time allocation and power allocation, subject to the average AoI requirement.
From (\ref{eq:Poutt}), (\ref{eq:Poutr}), and (\ref{eq:sumthP1}), the time-allocation parameters are  $\alpha _t$, $\alpha _r$, and $\alpha _{AP}$, and the power-allocation parameters are ${\beta _{t,TEP}}$ for ${U_t}$ and ${\beta _{r,TEP}}$ for ${U_r}$. From (\ref{eq:AoI}), the average AoI for the TEP scheme is ${\Delta _{TEP}} = \frac{1}{{{\Phi _{TEP}}}}$.

The optimization problem can be formulated as
\vspace{-0.6cm}
\begin{spacing}{1}
\begin{align}
&{{\cal P}_1}:\mathop {\max }\limits_{\begin{array}{*{20}{c}}
{{\alpha _t},{\alpha _r},{\alpha _{{\rm{AP}}}}}\\
{{\beta _{t,TEP}},{\beta _{r,TEP}}}
\end{array}}\!\!\!\!\!\!\! T^{TEP}({\alpha _t},{\alpha _r},{\alpha _{{\rm{AP}}}},{\beta _{t,TEP}},{\beta _{r,TEP}}) \label{YY}\\
&s.t.\ \ C1:0 < {\alpha _t},{\alpha _r},{\alpha _{{\rm{AP}}}} < 1,\tag{\ref{YY}{a}} \label{YYa}\\
&\ \ \ \ \ \ C2:{\alpha _t} + {\alpha _r} + {\alpha _{{\rm{AP}}}} = 1,\tag{\ref{YY}{b}} \label{YYb}\\
&\ \ \ \ \ \ C3:0 < {\beta _{t,TEP}},{\beta _{r,TEP}} < 1,\tag{\ref{YY}{c}} \label{YYc}\\
&\ \ \ \ \ \ C4:{\beta _{t,TEP}} + {\beta _{r,TEP}} = 1,\tag{\ref{YY}{d}} \label{YYd}\\
&\ \ \ \ \ \ C5:{\Delta _{TEP}} = \frac{1}{{{\Phi _{TEP}}}} < {\Delta _{th}}.\tag{\ref{YY}{e}} \label{YYe}
\vspace{-0.1cm}
\end{align}
\end{spacing}

Constraint (39a) specifies the range of ${\alpha _t}$, ${\alpha _r}$, and ${\alpha _{{\rm{AP}}}}$. Constraint (39b) illustrates the relationship between ${\alpha _t}$, ${\alpha _r}$ and ${\alpha _{{\rm{AP}}}}$. To simplify the analysis, let ${\alpha _t}$ and ${\alpha _r}$ be equal, i.e., ${\alpha _t} = {\alpha _r} = \frac{{\left( {1 - {\alpha _{{\rm{AP}}}}} \right)}}{2}$.
Constraint (39c) ensures the range of energy-splitting ratios of the STAR-RIS, i.e., the range of power-allocation ratios for each user.
Constraint (39d) is set to satisfy the law of energy conservation.
Constraint (39e) guarantees that the  average AoI is less than the preset  threshold ${\Delta _{th}}$.

\subsubsection{EEP}
For the EEP scheme, the joint time allocation and power allocation optimization problem, subject to average AoI, can be formulated as
\vspace{-0.6cm}
\begin{spacing}{1}
\begin{align}
&{{\cal P}_2}:\mathop {\max }\limits_{\begin{array}{*{20}{c}}
{{\alpha _{ET}},{\alpha _{IT}}}\\
{{\beta _{t,EEP}},{\beta _{r,EEP}}}
\end{array}}\!\!\!\!\!\!\!  T^{EEP}\left( {{\alpha _{ET}},{\alpha _{IT}},{\beta _{t,EEP}},{\beta _{r,EEP}}} \right)\label{XX}\\
&s.t. \ \ C1:0 < {\alpha _{ET}},{\alpha _{IT}} < 1,\tag{\ref{XX}{a}} \label{XXa}\\
&\ \ \ \  \ \ C2:{\alpha _{ET}} + {\alpha _{IT}} = 1,\tag{\ref{XX}{b}} \label{XXb}\\
&\ \ \ \  \ \ C3:0 < {\beta _{t,EEP}},{\beta _{r,EEP}} < 1,\tag{\ref{XX}{c}} \label{XXc}\\
&\ \ \ \  \ \ C4:{\beta _{t,EEP}} + {\beta _{r,EEP}} = 1,\tag{\ref{XX}{d}} \label{XXd}\\
&\ \ \ \  \ \ C5:{\Delta _{EEP}} = \frac{1}{{{\Phi _{EEP}}}} < {\Delta _{th}}.\tag{\ref{XX}{e}} \label{XXe}
\vspace{-0.1cm}
\end{align}
\end{spacing}

Constraints (40a) and (40b) specify the range and relationship between ${\alpha _{ET}}$ and ${\alpha _{IT}}$. Constraint (40c) ensures the range of
energy-splitting ratios of the STAR-RIS, i.e., the range of power-allocation ratios for each user. Constraint (40d) is set to satisfy
the law of energy conservation. Constraint (40e) guarantees that the average AoI is less than the preset threshold ${\Delta _{th}}$.

\subsection{Proposed Algorithm}
${{\cal P}_1}$ and ${{\cal P}_2}$ aim to maximize the sum throughput, which is related to the outage probability.
Since the outage probability expression is intractable, the derivative-free optimization method is considered to obtain the optimal solution. Moreover, as the scale of the network grows, it is difficult for the optimization problem to be solved in an acceptable amount of time using an exhaustive search method. Therefore, the genetic-algorithm (GA) based time allocation and power allocation (GA-TAPA) method is proposed to solve ${{\rm \mathcal{P} }_1}$ and ${{\cal P}_2}$.
The details of the proposed GA-TAPA method is summarized in Algorithm 1.

The GA is a meta-heuristic algorithm and an efficient global optimization method, which adopts the idea of survival of the fittest as its evolution principle to reach the optimal solution\cite{9802508,8490683}.
With the proposed GA-TAPA algorithm, which keeps the feature of GA, the
numerical optimized results can be obtained.
This derivative-free optimization is practical because it does not require the computation of gradients\cite{9691347}.

The time complexity of the proposed algorithm  is calculated.
Here,  binary coded GA is used. $\Xi$ is defined as the bit length of the binary code and chromosome
length.
We give the time complexity of each process and compute the overall complexity of the proposed algorithm.
According to the population size $\varepsilon$ and the chromosome length $\Xi$, the time complexity of initializing the population is given by $O(\varepsilon \times \Xi )$.
The time complexity of fitness evaluation for the maximum generation ${G_{en}}$ is $O(\varepsilon \times {G_{en}})$.
The time complexity of the selection operation, crossover operation, and mutation operation is $O({\varepsilon ^2} \times {G_{en}}) + O(\varepsilon  \times {G_{en}}) + O(\varepsilon  \times {G_{en}}) \approx O({\varepsilon ^2} \times {G_{en}})$.
Hence, the overall time complexity of the proposed GA-TAPA algorithm is approximated as $O({\varepsilon ^2} \times {G_{en}})$.
\vspace{-0.1cm}
\begin{algorithm}
\begin{small}
\begin{spacing}{1}
    \caption{Genetic-Algorithm Based Time Allocation and Power Allocation (GA-TAPA)}
    \label{alg:AOA}
    \renewcommand{\algorithmicrequire}{\textbf{Input:}}
    \renewcommand{\algorithmicensure}{\textbf{Output:}}
    \begin{algorithmic}[1]
        \REQUIRE  The size of the population $\varepsilon$, the iteration of GA-TAPA ${G_{en}}$, the parameters of GA ${q_t}$, ${p_t}$, ${p_m}$. 
        \ENSURE ${\alpha_t}$, ${\alpha_r}$, ${\alpha_{AP}}$, ${\beta _{t,{TEP}}}$, ${\beta _{r,{TEP}}}$ for the TEP scheme or ${\alpha_{ET}}$, ${\alpha_{IT}}$, ${\beta _{t,{EEP}}}$, ${\beta _{r,{EEP}}}$ for the EEP scheme.
        \STATE Randomly initialize the population based on $\varepsilon$, (39a), (39b), (39c), (39d) for the TEP, or $\varepsilon$, (40a), (40b), (40c), (40d) for the EEP.
        \FOR{ $i = 1:{G_{en}}$}{
        \STATE Evaluate the population, i.e., calculate the penalty fitness value of the individuals according to (39) and (39e) for the TEP scheme or (40) and (40e) for the EEP scheme.
        \STATE Rank the individuals according to their penalty fitness values. Select the father generations with the selection probability ${q_t}$ and randomly select mother generation.
        \STATE Crossover the father and mother generations with the crossover
         probability ${p_t}$ to generate new population.
         \STATE For each individual, do the mutation operation with
         the mutation probability ${p_m}$.
         \STATE Update population.
         }
         \ENDFOR
         \STATE Calculate the fitness value of the population and find the best fitness individual.
    \end{algorithmic}
    \end{spacing}
    \end{small}
\end{algorithm}

\vspace{-1.0cm}
\section{Results and Discussion} 	
In this section, the outage probability, sum throughput and average AoI of the proposed STAR-RIS aided wireless-powered NOMA system are evaluated, where the TEP and EEP are considered.
In the simulations, ${P_{{\rm{AP}}}} = 1$ Watt (W). The parameters of the path loss are set as follows: ${d_0} = 30$~m, ${d_r} = 4$~m, ${d_t} = 2$~m, and
${\vartheta _0} = {\vartheta _\chi } = 2$.
Moreover, unless stated otherwise, it is assumed that
${\beta _{r,{TEP}}} = {\beta _{r,{EEP}}} = 0.4$, ${\beta _{t,{TEP}}} = {\beta _{t,{EEP}}} = 0.6$, ${\alpha _t} = {\alpha _r} = 0.25$, ${\alpha _{AP}} = 0.5$, ${\alpha _{ET}} = 0.5$, and ${\alpha _{IT}} = 0.5$.
Regarding the Nakagami-$m$ fading channel parameters, we assume ${m_i} = {m_{t,i}} = {m_{r,i}} = 2$ and ${\Omega _i} = {\Omega _{t,i}} = {\Omega _{r,i}} = 1$.
To demonstrate the advantages of the proposed system, conventional RIS (C-RIS) schemes are considered as baseline schemes for comparison purposes.
For one C-RIS scheme, a reflecting-only RIS and a transmitting-only RIS at the same location of the STAR-RIS are deployed to achieve full-space coverage, where each reflecting/transmitting-only RIS is equipped with $\frac{N}{2}$ elements \cite{9834288}. Furthermore, both C-RIS wireless-powered NOMA (C-RIS-NOMA) and C-RIS wireless-powered TDMA (C-RIS-TDMA) schemes are considered. In addition, the STAR-RIS aided wireless-powered TDMA system is referred to as STAR-RIS-TDMA.
\vspace{-1cm}
\begin{figure}[!htb]
\begin{tabular}{cc}
\begin{minipage}[t]{0.48\linewidth}
    \includegraphics[width = 1\linewidth]{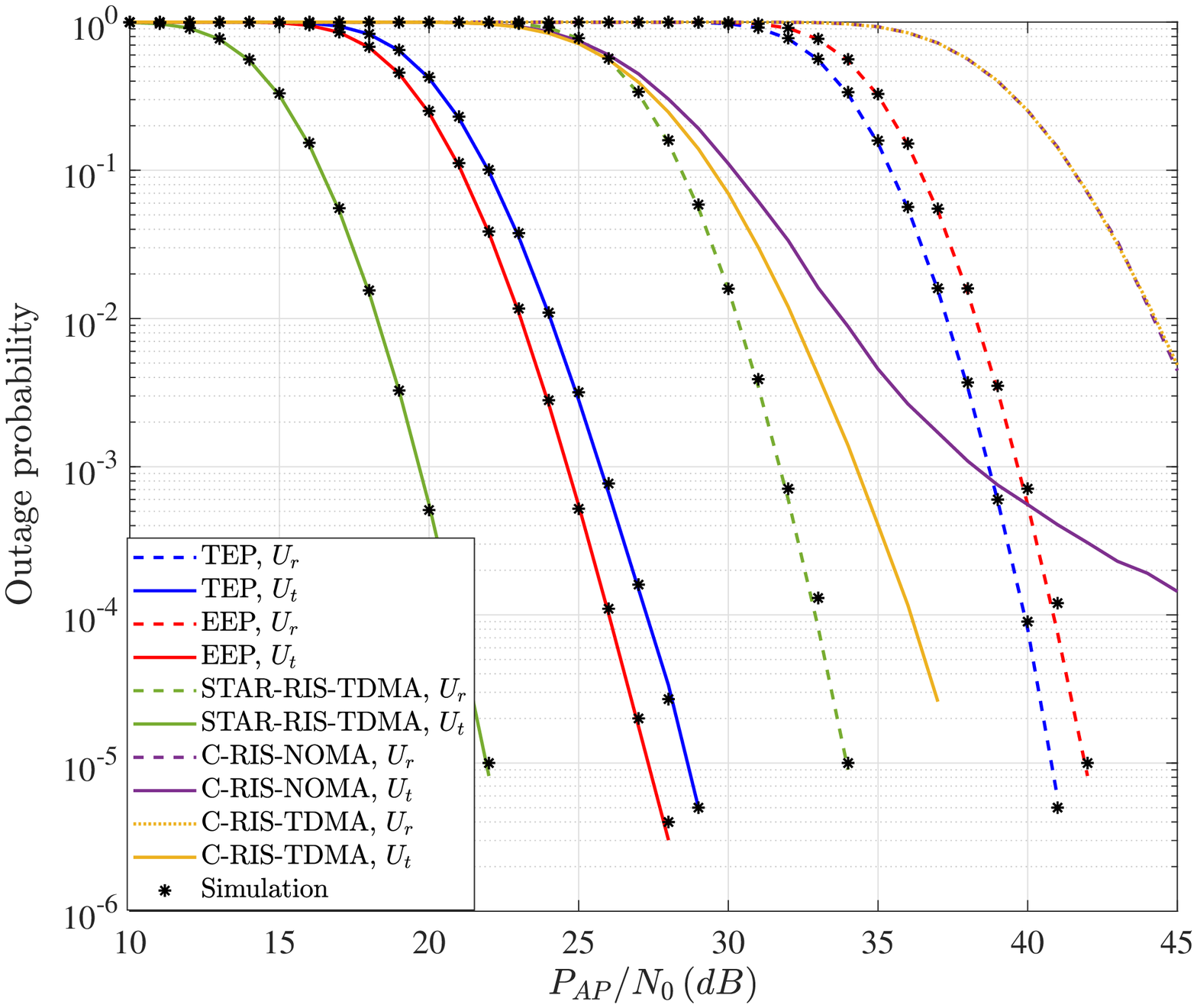}
    \vspace{-1cm}
    \captionsetup{font={small}, justification=raggedright}
    \caption{Outage probability versus transmit SNR of various systems, where $N = 30$ and $R =1$.}
    \label{fig:PoutVS}
\end{minipage}
\begin{minipage}[t]{0.48\linewidth}
    \includegraphics[width = 1\linewidth]{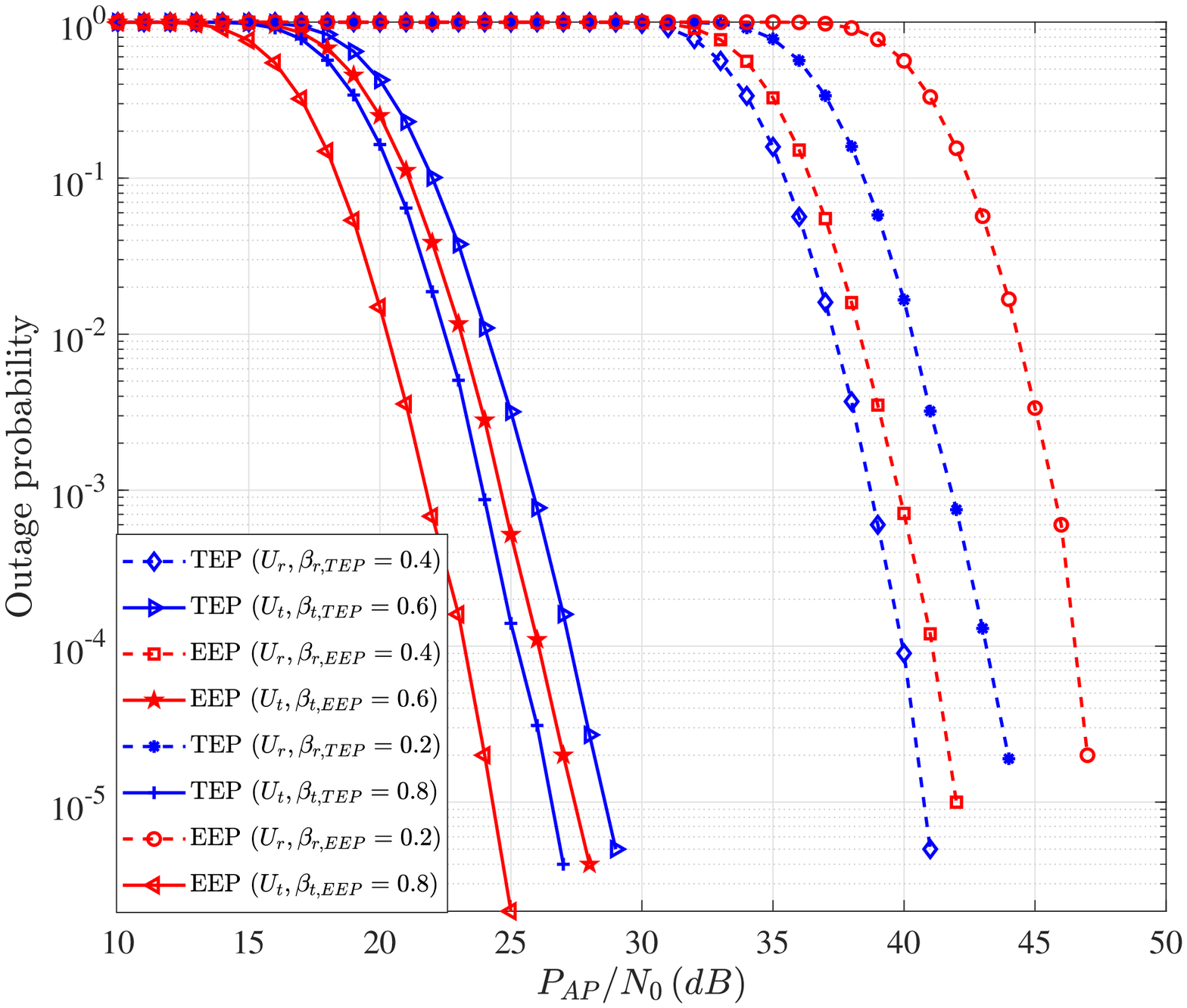}
    \vspace{-1cm}
    \captionsetup{font={small}, justification=raggedright}
    \caption{Outage probability of the proposed system with the TEP and EEP schemes for different power allocation coefficients, where $N = 30$ and $R =1$.}
    \label{fig:PoutVSbeta}
\end{minipage}
\end{tabular}
\vspace{-1cm}
\end{figure}

Fig.~\ref{fig:PoutVS} shows the outage probability versus transmit SNR of various systems, where $N =30$ and $R =1$.
In this figure, theoretical results are consistent with the simulation
results, which validates the proposed analyses.
It can be observed that the proposed system shows better outage probability performance than the C-RIS-NOMA and C-RIS-TDMA systems. The reason behind that  is that STAR-RIS can configure full-space electromagnetic propagation environments while the conventional RIS needs a reflecting-only RIS and a transmitting-only RIS to achieve full-space coverage. Compared to the C-RIS, the STAR-RIS has more adjustable parameters to adjust the channel conditions of users.
Moreover, it can be seen that the outage probability performance of the STAR-RIS-TDMA scheme is better than that of the proposed scheme, because inter-user interference occurs in the proposed scheme.
Furthermore, it can be seen that for $U_t$, the EEP scheme shows better performance than the TEP scheme; while for $U_r$, the TEP scheme performs better compared to the EEP scheme.
This is because the EEP scheme allocates more resources to $U_t$, while the TEP scheme balances the energy harvested by the two users in the downlink to reduce the performance gap.

Fig.~\ref{fig:PoutVSbeta} depicts the outage probability of the proposed system with the TEP and EEP schemes for different power allocation coefficients.
It is observed that $U_r$ is more sensitive to changes in the power allocation compared to $U_t$.
For example, for the EEP scheme, an increment of $0.2$ in the power allocation results in a $6$~dB gain for $U_r$ while $U_t$ can only achieve a $3$~dB gain at an outage probability of ${10^{ - 4}}$.
This is because $U_r$ is a far-side user, where more power allocation helps $U_r$ significantly improve its performance and reduce the double near-far impact, which indicates that the far user harvests less energy, but needs  more  energy to transmit information during the uplink.
Moreover, it can be seen that the EEP scheme is more sensitive to changes in the power allocation compared to the TEP scheme due to the energy-splitting protocol in both uplink and downlink.
\vspace{-1cm}
\begin{figure}[!htb]
\begin{tabular}{cc}
\begin{minipage}[t]{0.48\linewidth}
    \includegraphics[width = 1\linewidth]{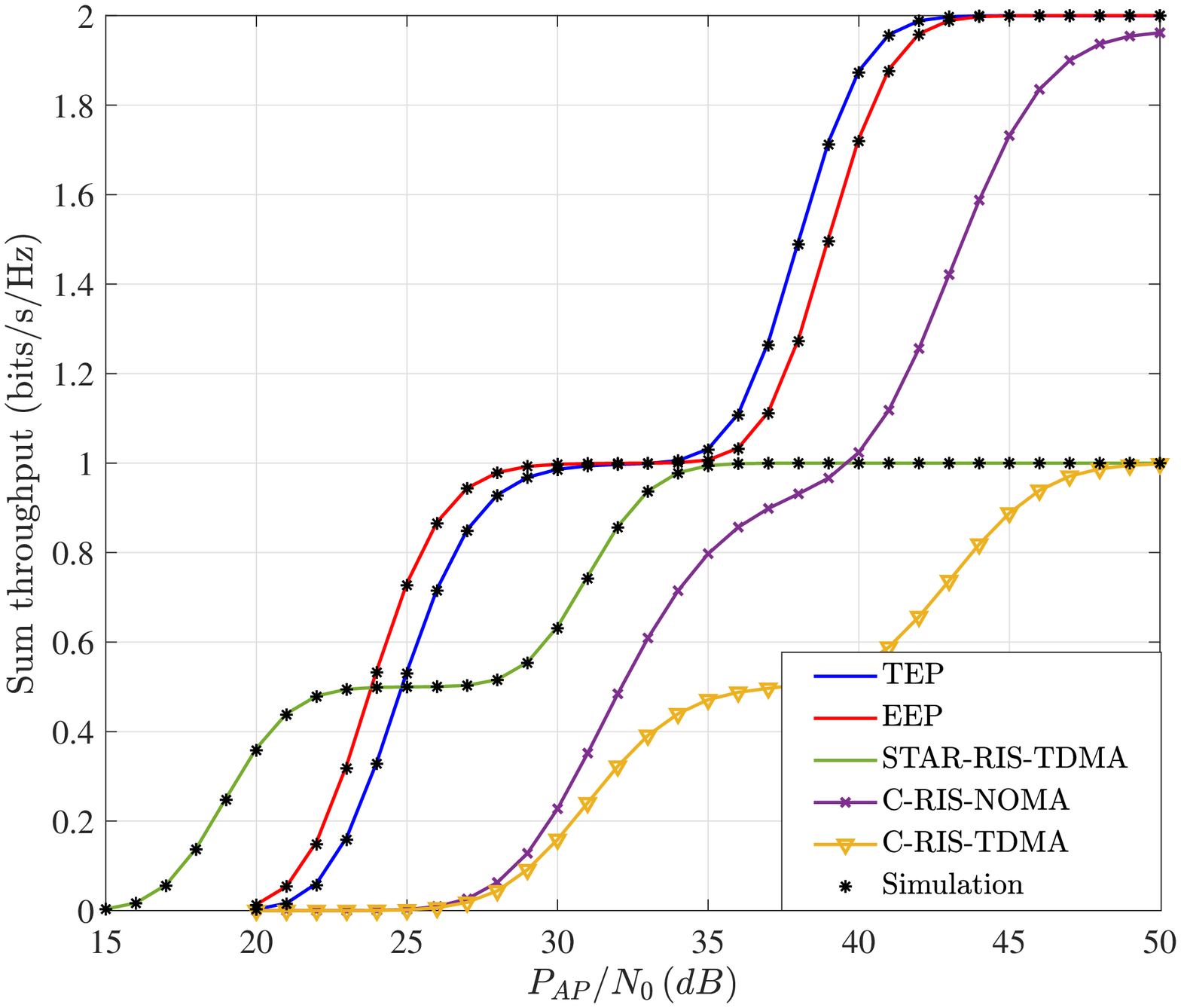}
    \vspace{-1cm}
    \captionsetup{font={small}, justification=raggedright}
    \caption{Sum throughput versus transmit SNR for \\various systems, where $N =30$ and $R =2$.}
    \label{fig:SumthVS}
\end{minipage}
\begin{minipage}[t]{0.48\linewidth}
    \includegraphics[width = 1\linewidth]{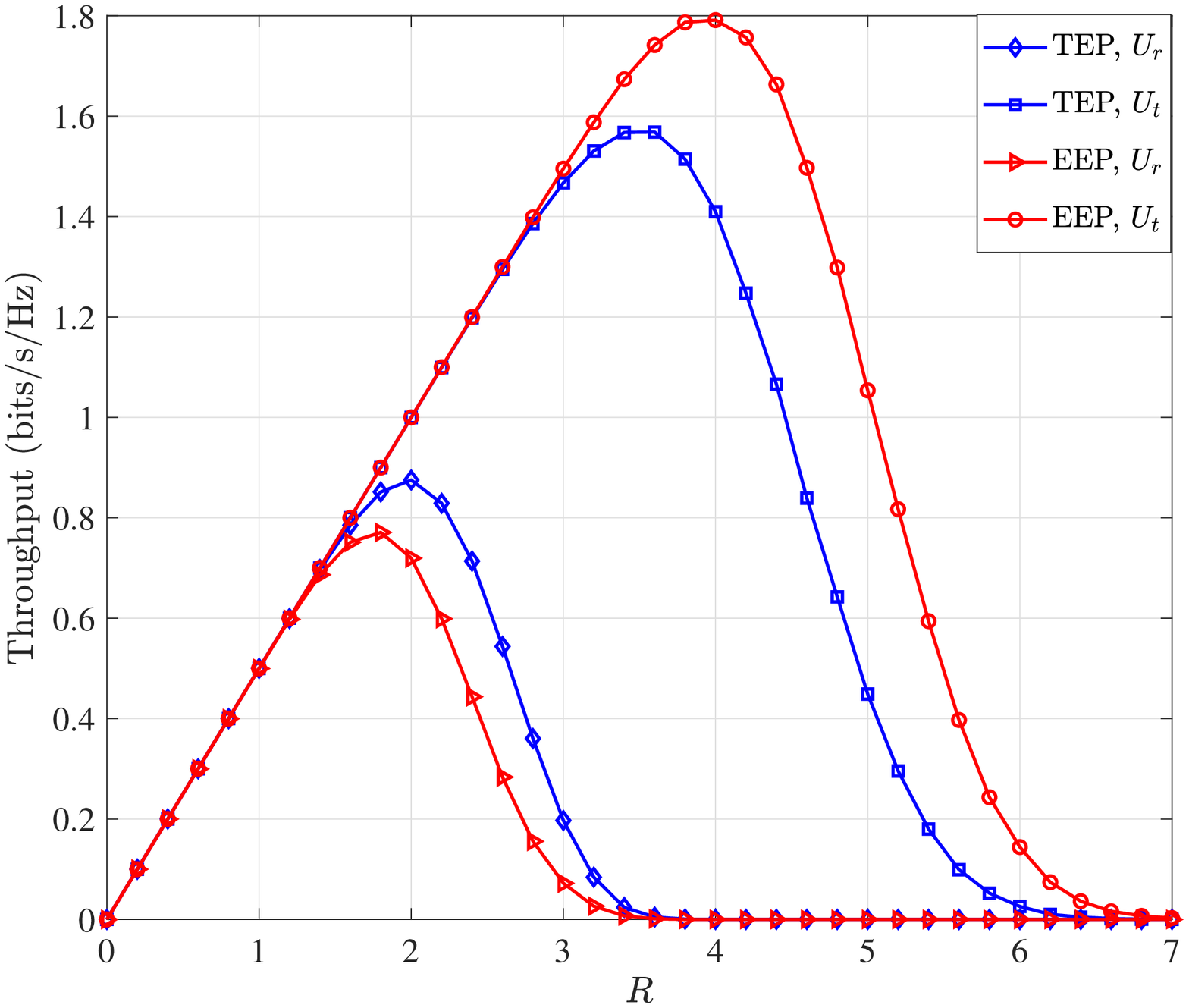}
    \vspace{-1cm}
    \captionsetup{font={small}, justification=raggedright}
    \caption{Throughput versus $R$ for the proposed system with the TEP and EEP schemes, where $N = 30$ and $\frac{{{P_{AP}}}}{{{N_0}}}=40$~dB.}
    \label{fig:thVSR}
\end{minipage}
\end{tabular}
\vspace{-1cm}
\end{figure}

Fig.~\ref{fig:SumthVS} illustrates the sum throughput versus transmit SNR of various systems, where $N =30$ and $R =2$.
It can be observed that the proposed system shows better sum throughput performance than the baseline and STAR-RIS-TDMA schemes.
The reason behind this is that NOMA allows $U_t$ and $U_r$ to use the same time-frequency resource block,
thus achieving a multiplexing gain compared to TDMA, and NOMA allows the STAR-RIS to exploit the proper power allocation to achieve better performance.
At low SNR, the performance of STAR-RIS-TDMA is a slightly better than that of the proposed system with the TEP and EEP schemes. This is because the TEP and EEP schemes use the energy-splitting protocol in the uplink, thus causing opposite-side leakage, which indicates that some of the uplink signals of $U_r$ are transmitted and some of the uplink signals of $U_t$ are reflected.
Moreover, it can be seen that there is an intersection point between the C-RIS-NOMA and STAR-RIS-TDMA curves.
This is because at low SNR, the beamforming gain of the STAR-RIS is more prominent while at high SNR, the multiplexing gain increases and eventually exceeds the beamforming gain to become the dominant factor\cite{9834288}.
Furthermore, it can be seen that the sum throughput performance of the EEP scheme is better than that of the TEP scheme at low SNR while the TEP outperforms the EEP at high SNR.
The reason behind this is that the outage probability of the EEP scheme is better at low SNR while the outage probability of the TEP scheme is better at high SNR.

Fig.~\ref{fig:thVSR} plots the throughput versus $R$ for the proposed system with the TEP and EEP schemes, where $N = 30$ and $\frac{{{P_{AP}}}}{{{N_0}}}=40$~dB.
It can be observed that there exists an optimal value for $R$ that maximizes the throughput.
The reason behind this is that at low rates, the outage probability is low and the throughput is limited by $R$. In contrast, at high rates, the throughput is limited by the high outage probability.
Hence, an optimal value of $R$ that maximizes the throughput must exist.
Moreover, it can be seen that for $U_r$ the throughput performance of the TEP scheme is better than that of the EEP scheme for small values of $R$, while for $U_t$ the EEP scheme outperforms the TEP scheme for large values of $R$.

\begin{figure}[!htb]
\begin{tabular}{cc}
\begin{minipage}[t]{0.48\linewidth}
    \includegraphics[width = 1\linewidth]{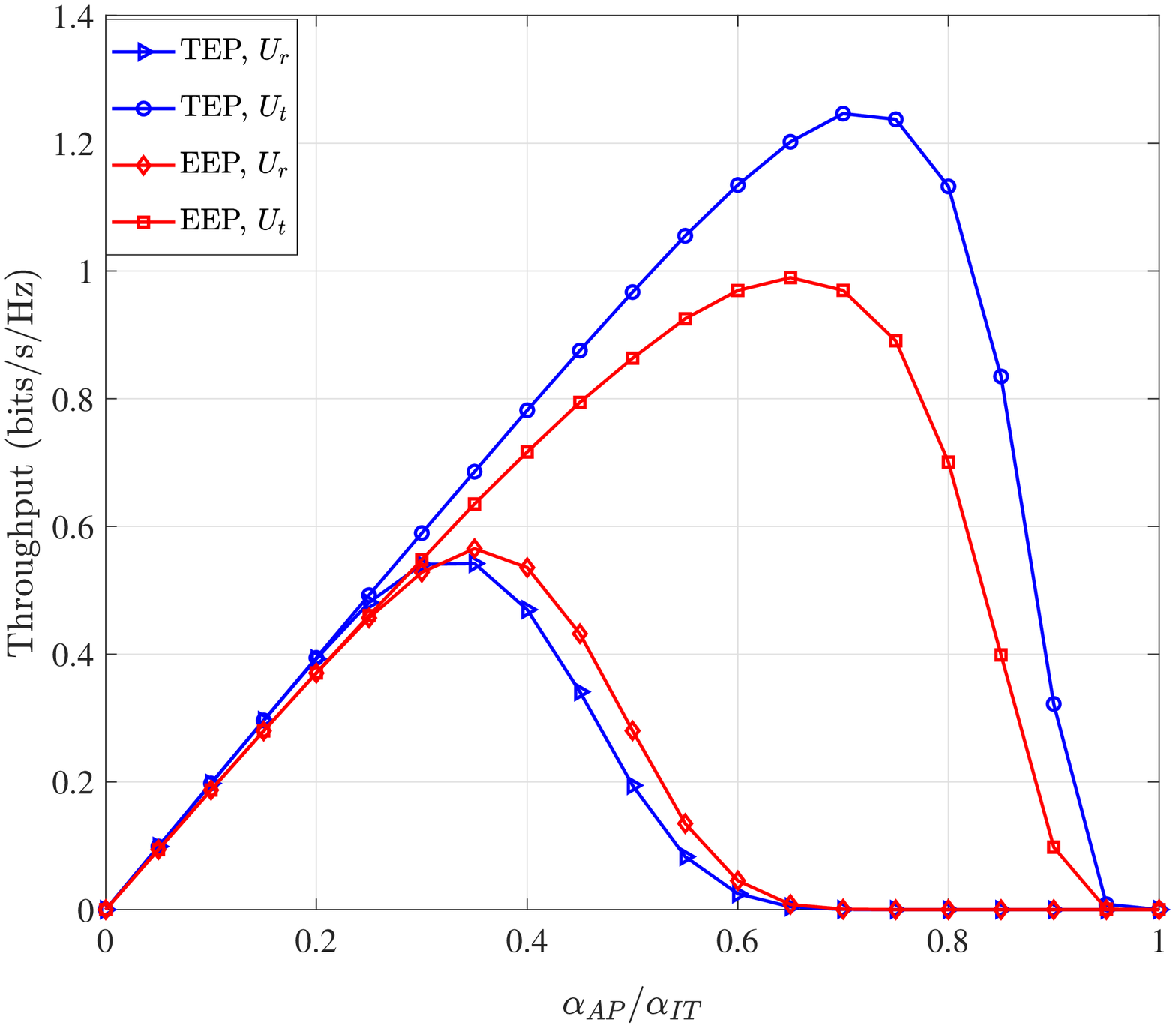}
    \vspace{-1cm}
    \captionsetup{font={small}, justification=raggedright}
    \caption{Throughput versus ${\alpha _{AP}}/{\alpha _{IT}}$ for the proposed system with the TEP and EEP schemes, where \\$N = 30$, $\frac{{{P_{AP}}}}{{{N_0}}}=35$~dB, and $R = 2$.}
    \label{fig:thVStime}
\end{minipage}
\begin{minipage}[t]{0.48\linewidth}
    \includegraphics[width = 1\linewidth]{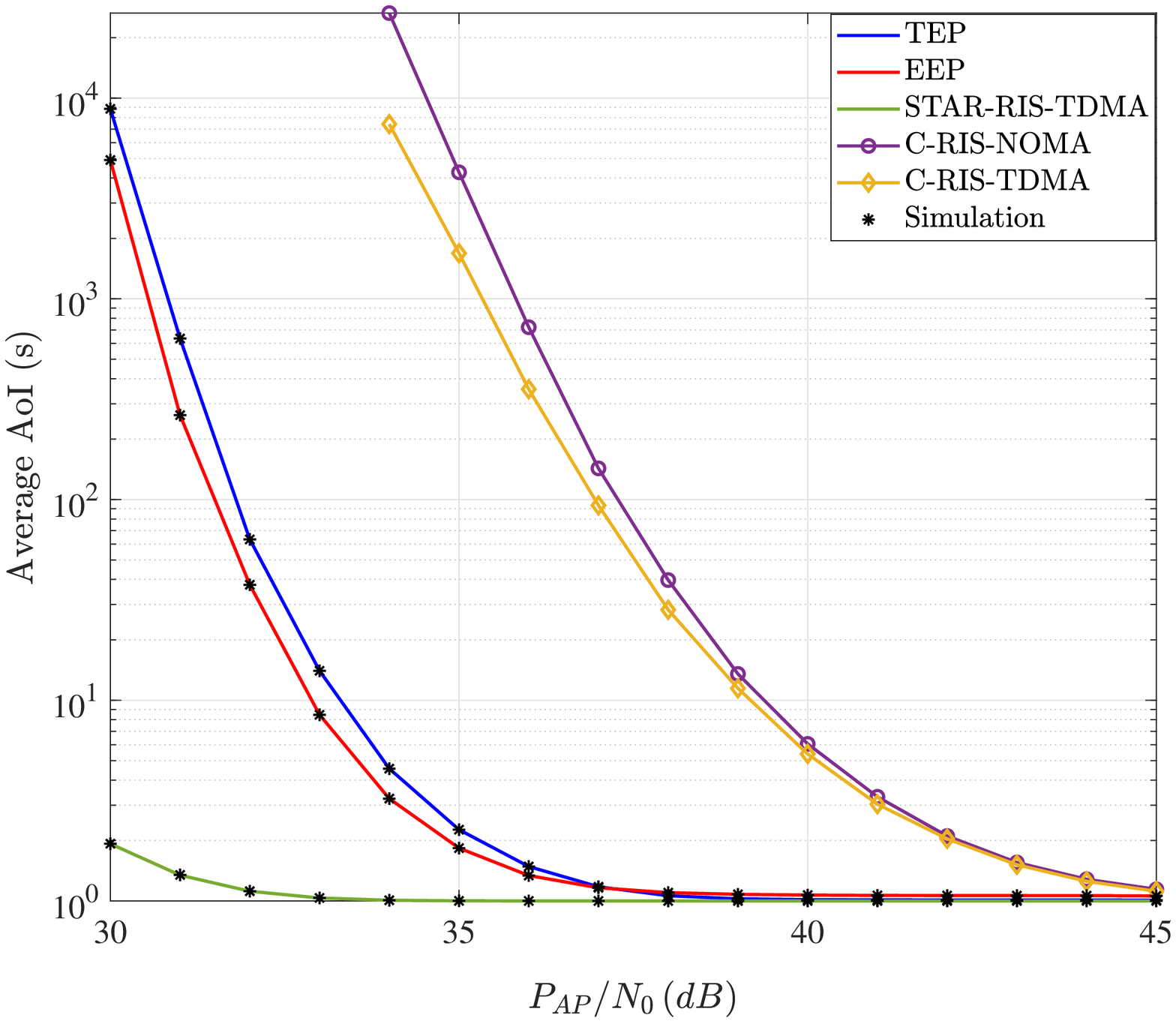}
    \vspace{-1cm}
    \captionsetup{font={small}, justification=raggedright}
    \caption{Average AoI versus transmit SNR of various systems, where $N = 32$ and $R = 2$.}
    \label{fig:AoIVS}
\end{minipage}
\end{tabular}
\vspace{-1.2cm}
\end{figure}

Fig.~\ref{fig:thVStime} shows the throughput versus ${\alpha _{AP}}/{\alpha _{IT}}$ of the proposed system with the TEP and EEP schemes, where $N = 30$, $\frac{{{P_{AP}}}}{{{N_0}}}=35$~dB, $R = 2$, ${\beta _{r,{TEP}}} = {\beta _{r,{EEP}}} = 0.6$, and ${\beta _{t,{TEP}}} = {\beta _{t,{EEP}}} = 0.4$.
It can be seen that the throughput increases first and then decreases with ${\alpha _{AP}}/{\alpha _{IT}}$.
The reason lies in that there is a tradeoff between the ET and IT phases. For longer IT phases, more data can be transferred while the users harvest less energy and the transmission power is reduced.
Moreover, it can be observed that the optimal duration for the IT phase is smaller for the far-side user $U_r$, compared to the near-side user $U_t$, since $U_r$ needs more time in the ET phase to harvest energy.
Furthermore, since the EEP scheme allocates more power to $U_r$, for $U_r$ the EEP scheme performs better than the TEP scheme while for $U_t$ the TEP scheme  outperforms the EEP scheme.
\vspace{-0.8cm}

Fig.~\ref{fig:AoIVS} plots the average AoI versus transmit SNR of various systems, where $N = 32$, $R = 2$, ${\beta _{r,{TEP}}} = {\beta _{r,{EEP}}} = 0.6$, and ${\beta _{t,{TEP}}} = {\beta _{t,{EEP}}} = 0.4$.
It is observed that the performance of the STAR-RIS-TDMA scheme is better than that of the proposed system, since the STAR-RIS-TDMA scheme has no inter-user interference and better success probability.
Moreover, it can be seen that there is an intersection between the TEP and EEP schemes.
The average AoI of the EEP scheme is smaller than that of the TEP scheme at low SNRs, while the TEP scheme shows a better average AoI performance at high SNRs.
This is because for low SNRs, the EEP scheme can allocate more power to ${U_r}$ to reduce the performance gap between the two users, thus increasing the success probability.
For high SNRs, the EEP scheme allocates excessive power to ${U_r}$, which deteriorates the performance in terms of average AoI.

\begin{figure}[!htb]
\begin{tabular}{cc}
\begin{minipage}[t]{0.48\linewidth}
    \includegraphics[width = 1\linewidth]{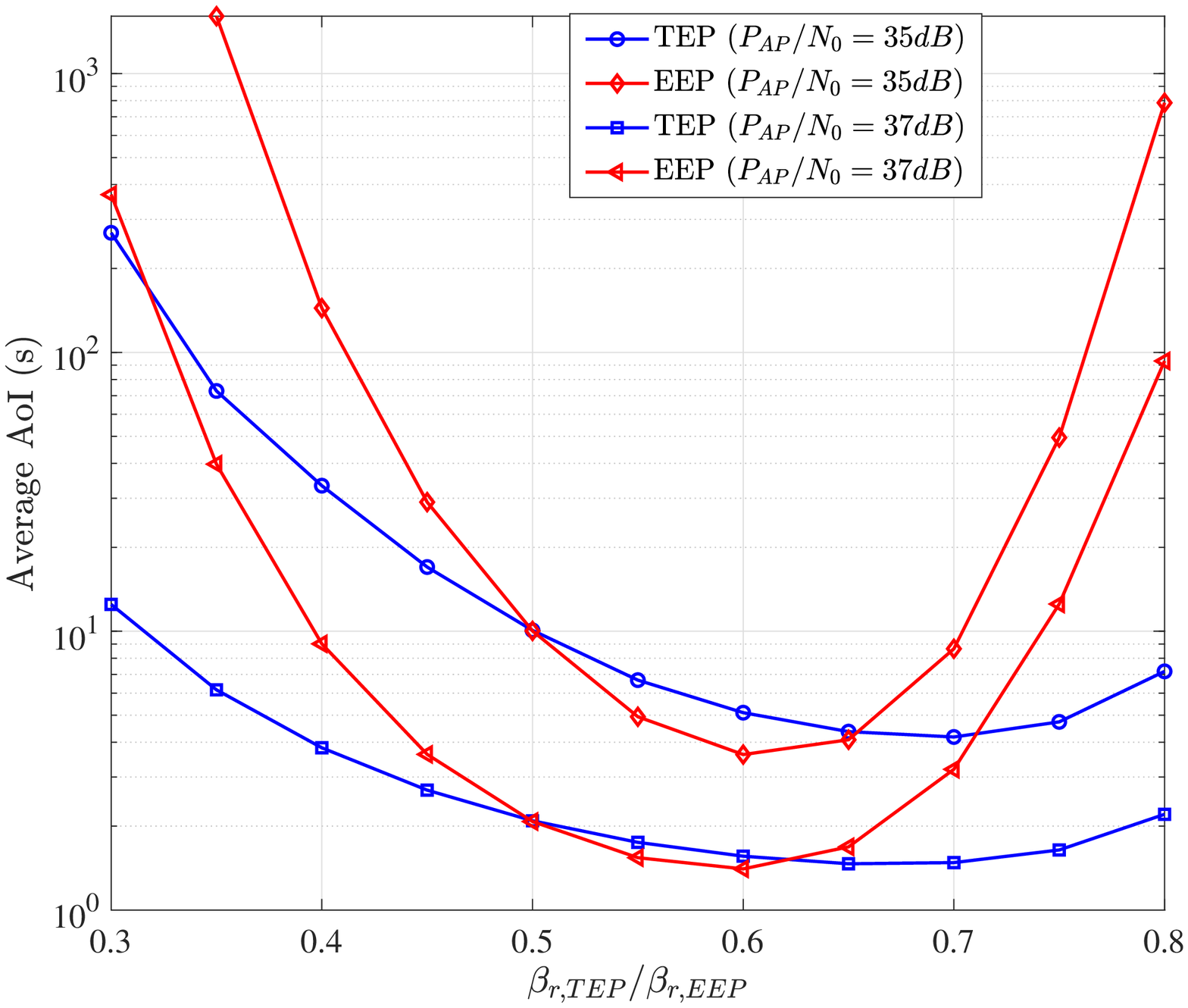}
    \vspace{-0.8cm}
    \captionsetup{font={small}, justification=raggedright}
    \caption{Average AoI versus ${\beta _{r,TEP}}/{\beta _{r,EEP}}$ for the proposed system with the TEP and EEP schemes, where $N = 30$ and $R = 2$.}
    \label{fig:AoIVSbeta}
\end{minipage}
\begin{minipage}[t]{0.48\linewidth}
    \includegraphics[width = 1\linewidth]{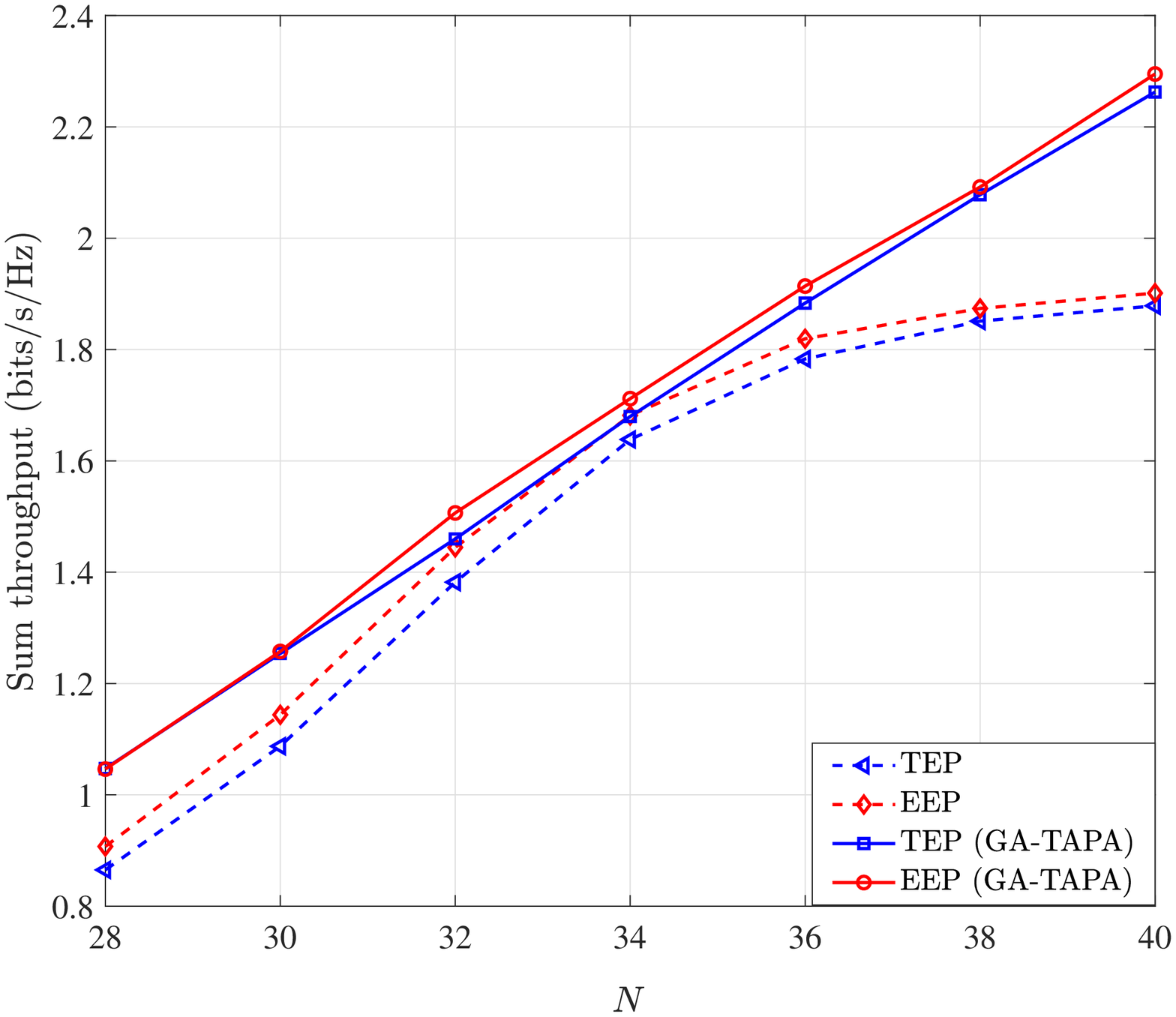}
    \vspace{-0.8cm}
    \captionsetup{font={small}, justification=raggedright}
    \caption{Optimized sum throughput versus the number of STAR-RIS $N$ for the proposed system with the EEP and TEP schemes, where $\frac{{{P_{AP}}}}{{{N_0}}}=35$~dB and $R=2$.}
    \label{fig:youhua}
\end{minipage}
\end{tabular}
\vspace{-1cm}
\end{figure}

Fig.~\ref{fig:AoIVSbeta} illustrates the average AoI versus energy-splitting ratio ${\beta _{r,TEP}}/{\beta _{r,EEP}}$ for the proposed system with the TEP and EEP schemes, where $N = 30$ and $R = 2$.
It can be observed that for the EEP scheme, the optimal value of ${\beta _{r,EEP}}$ is about $0.6$. The reason behind this is that low values of ${\beta _{r,EEP}}$ cause deteriorations in the average AoI due to poor performance for ${U_r}$, and high values of ${\beta _{r,EEP}}$ allocate excessive power to ${U_r}$,
thus resulting in 
serious inter-user interference, which is not conducive to SIC decoding.
Moreover, it can be seen that the TEP scheme yields better performance most of the time.
This is because the TEP scheme uses the  time-switching protocol in the downlink, which balances the resource allocation and improves the performance of ${U_r}$ for low values of ${\beta _{r,TEP}}$, and avoids excessive power allocation for high values of ${\beta _{r,TEP}}$.
\vspace{-0.7cm}

Fig.~\ref{fig:youhua} shows the optimized sum throughput versus the number of the STAR-RIS $N$ for the proposed system with EEP and TEP schemes, where $\frac{{{P_{AP}}}}{{{N_0}}}=35$~dB and $R=2$.
It can be observed that our proposed GA-TAPA method outperforms the baseline TEP and EEP schemes.
The GA-TAPA method can adjust the parameters to improve the sum throughput performance according to variations in $N$.
For example, the TEP scheme with GA-TAPA method can achieve $1$~dB gain at sum throughput of $1.2$ compared to the TEP scheme.
Moreover, it can be seen that when $N$ becomes large, the growth in the sum throughput of the TEP and EEP schemes becomes smaller and tends to saturation while the TEP and EEP schemes with GA-TAPA method can adjust parameters and continue to improve the performance in terms of sum throughput.

\section{Conclusion} \label{sect:conclusion}
In this paper, a STAR-RIS aided wireless-powered NOMA system was proposed, where the TEP and EEP schemes are considered.
The outage probability, sum throughput, and average AoI of the proposed system with the TEP and EEP schemes and the STAR-RIS aided wireless-powered TDMA system were derived over Nakagami-$m$ fading channels.
Simulation and numerical results show that the proposed system with the TEP and EEP schemes not only yields better performance than baseline schemes, but also offers higer sum throughputs at the cost of the outage probability and average AoI compared to the STAR-RIS aided wireless-powered TDMA system.
Moreover, the GA-TAPA algorithm has been designed to optimize the sum throughput while satisfying an average AoI constraint by jointly optimizing the time-allocation and power-allocation parameters.
Results show that the proposed GA-TAPA method can significantly improve the sum throughput. Thanks to these advantages, the proposed STAR-RIS aided wireless-powered NOMA system can provide a new way to solve the multi-user interference and energy supply problems in the future IoT systems.

\appendices
\section{}
Here, the $n$-th moment of ${H_{{h_i}{g_{t,i}}}}$ is derived.
Since ${h_i}\sim {\rm{Nakagami}}\left( {{m_i},{\Omega _i}} \right)$ and ${g_{t,i}}\sim {\rm{Nakagami}}\left( {{m_{t,i}},{\Omega _{t,i}}} \right)$, the PDF of ${h_i}$ and ${g_{t,i}}$ are given by
\vspace{-0.3cm}
\begin{spacing}{1}
\begin{small}
\begin{eqnarray}
{f_{{h_i}}}\left( {x;{m_i},{\Omega _i}} \right) \!=\! \frac{{2{m_i}^{{m_i}}}}{{\Gamma \left( {{m_i}} \right){\Omega _i}^{{m_i}}}}{x^{2{m_i} - 1}}{e^{ - \frac{{{m_i}}}{{{\Omega _i}}}{x^2}}},{f_{{g_{t,i}}}}\left( {x;{m_{t,i}},{\Omega _{t,i}}} \right) \!=\! \frac{{2{m_{t,i}}^{{m_{t,i}}}}}{{\Gamma \left( {{m_{t,i}}} \right){\Omega _{t,i}}^{{m_{t,i}}}}}{x^{2{m_{t,i}} - 1}}{e^{ - \frac{{{m_{t,i}}}}{{{\Omega _{t,i}}}}{x^2}}}\!\!\!.
\vspace{-0.7cm}
\end{eqnarray}
\end{small}
\end{spacing}

Considering $Z = XY$, ${f_Z}\left( z \right) = \int\limits_0^\infty  {\frac{1}{x}{f_Y}\left( {\frac{z}{x}} \right)} {f_X}\left( x \right)dx$.
For ${H_{{h_i}{g_{t,i}}}} = {h_i}{g_{t,i}}$, the PDF of ${H_{{h_i}{g_{t,i}}}}$ is expressed as
\vspace{-0.5cm}
\begin{spacing}{1}
\begin{small}
\begin{eqnarray}
\label{eq:fH}
{f_{{H_{{h_i}{g_{t,i}}}}}}\left( z \right) &= \frac{{4{m_i}^{{m_i}}{m_{t,i}}^{{m_{t,i}}}}}{{\Gamma \left( {{m_i}} \right){\Omega _i}^{{m_i}}\Gamma \left( {{m_{t,i}}} \right){\Omega _{t,i}}^{{m_{t,i}}}}}{z^{2{m_{t,i}} - 1}}
  \int\limits_0^\infty  {{x^{2{m_i} - 2{m_{t,i}} - 1}}} {e^{\left( { - \frac{{{m_i}}}{{{\Omega _i}}}{x^2} - \frac{{{m_{t,i}}}}{{{\Omega _{t,i}}}}{{\left( {\frac{z}{x}} \right)}^2}} \right)}}dx.
  \vspace{-0.7cm}
\end{eqnarray}
\end{small}
\end{spacing}

According to (\!\!\cite{Gradshte1965Table}, eq.(3.478.4)), (\ref{eq:fH}) is reformulated as
\vspace{-0.3cm}
\begin{small}
\begin{equation}
\label{eq:fHH}
{f_{{H_{{h_i}{g_{t,i}}}}}}\left( z \right) = \frac{{4{\lambda _{{h_i}{g_{t,i}}}}^{{m_i} + {m_{t,i}}}}}{{\Gamma \left( {{m_i}} \right)\Gamma \left( {{m_{t,i}}} \right)}}{z^{{m_i} + {m_{t,i}} - 1}}{K_{{m_i} - {m_{t,i}}}}\left( {2z{\lambda _{{h_i}{g_{t,i}}}}} \right),
\vspace{-0.3cm}
\end{equation}
\end{small}
where ${\lambda _{{h_i}{g_{t,i}}}} = \sqrt {\frac{{{m_i}}}{{{\Omega _i}}}\frac{{{m_{t,i}}}}{{{\Omega _{t,i}}}}}$.

The $n$-th moment of ${H_{{h_i}{g_{t,i}}}}$ is defined as ${\mu _{{H_{{h_i}{g_{t,i}}}}}}\left( n \right) \buildrel \Delta \over = E\left( {H_{{h_i}{g_{t,i}}}^n} \right) = \int\limits_0^\infty  {{z^n}} {f_{{H_{{h_i}{g_{t,i}}}}}}\left( z \right)dz$. Using (\ref{eq:fHH}) and after some mathematical manipulations, ${\mu _{{H_{{h_i}{g_{t,i}}}}}}\left( n \right)$ is obtained as
\vspace{-0.1cm}
\begin{spacing}{1}
\begin{small}
\begin{equation}
{\mu _{{H_{{h_i}{g_{t,i}}}}}}\left( n \right) = \lambda _{{h_i}{g_{t,i}}}^{ - n}\frac{{\Gamma \left( {{m_i} + n/2} \right)\Gamma \left( {{m_{t,i}} + n/2} \right)}}{{\Gamma \left( {{m_i}} \right)\Gamma \left( {{m_{t,i}}} \right)}}.
\vspace{-0.5cm}
\end{equation}
\end{small}
\end{spacing}

\vspace{-0.5cm}
\section{}
Here, (\ref{eq:Poutt}) is derived.
Let $A = \frac{{{P_{AP}}l_t^2{\beta _{t,TEP}}{\alpha _t}}}{{{\alpha _{AP}}{N_0}}}$, $B = \frac{{{P_{AP}}l_r^2{\beta _{r,TEP}}{\alpha _r}}}{{{\alpha _{AP}}{N_0}}}$,  the first part of (\ref{eq:Poutt}), termed as $P_1$, can be written as
\vspace{-0.5cm}
\begin{spacing}{1}
\begin{small}
\begin{align}
\label{eq:P1}
{P_1} &\!=\! \Pr \left( \!{\frac{{{\gamma _{t,{\rm{TEP}}}}}}{{{\gamma _{r,{\rm{TEP}}}} + 1}} < {\gamma _{th}},\frac{{{\gamma _{r,{\rm{TEP}}}}}}{{{\gamma _{t,{\rm{TEP}}}} + 1}} < {\gamma _{th}}}\! \right)
 \!=\! \Pr \left(\! {\frac{{A{{\left| {{G_{{h_i}{g_{t,i}}}}} \right|}^4}}}{{B{{\left| {{G_{{h_i}{g_{r,i}}}}} \right|}^4} + 1}} < {\gamma _{th}},\frac{{B{{\left| {{G_{{h_i}{g_{r,i}}}}} \right|}^4}}}{{A{{\left| {{G_{{h_i}{g_{t,i}}}}} \right|}^4} + 1}} < {\gamma _{th}}}\! \right)\nonumber\\
 &= \Pr \left( \!\!{\frac{{B{{\left| {{G_{{h_i}{g_{r,i}}}}} \right|}^4}}}{{A{\gamma _{th}}}} - \frac{1}{A} \!<\! {{\left| {{G_{{h_i}{g_{t,i}}}}} \right|}^4} \!<\! \frac{{{\gamma _{th}}}}{A}\left( {B{{\left| {{G_{{h_i}{g_{r,i}}}}} \right|}^4} + 1} \right)}\!\! \right)\nonumber\\
 &= F\left( {\frac{{{\gamma _{th}}}}{A}\left( {B{{\left| {{G_{{h_i}{g_{r,i}}}}} \right|}^4} + 1} \right)} \right) - F\left( {\frac{{B{{\left| {{G_{{h_i}{g_{r,i}}}}} \right|}^4}}}{{A{\gamma _{th}}}} - \frac{1}{A}} \right)\nonumber\\
 &= \int\limits_0^\infty \!\! {F\left( {\frac{{{\gamma _{th}}}}{A}\left( {Bx + 1} \right)} \right)} f\left( x \right)dx -\!\!\! \int\limits_{\frac{{{\gamma _{th}}}}{B}}^\infty \!\!\! {F\left( {\frac{{Bx}}{{A{\gamma _{th}}}} - \frac{1}{A}} \right)} f\left( x \right)dx.
 \vspace{-0.3cm}
\end{align}
\end{small}
\end{spacing}
The first integral in (\ref{eq:P1}) can be calculated as
\vspace{-0.3cm}
\begin{spacing}{1}
\begin{small}
\begin{align}
\label{eq:P11}
&\int\limits_0^\infty  {F\left( {\frac{{{\gamma _{th}}}}{A}\left( {Bx + 1} \right)} \right)} f\left( x \right)dx = \int\limits_0^\infty  {\frac{1}{{\Gamma \left( {Nk} \right)}}\gamma \left( {Nk,\theta {{\left( {\frac{{{\gamma _{th}}}}{A}\left( {Bx + 1} \right)} \right)}^{\frac{1}{4}}}} \right)} \frac{{{\theta ^{Nk}}{e^{ - \theta {x^{\frac{1}{4}}}}}{x^{\frac{{Nk - 4}}{4}}}}}{{4\left( {Nk - 1} \right)!}}dx\nonumber\\
 &= 1 - \sum\limits_{m = 0}^{Nk - 1} {\frac{{{\theta ^{Nk + m}}}}{{4m!\left( {Nk - 1} \right)!}}} \int\limits_0^\infty  {{e^{ - \theta {{\left( {\frac{{{\gamma _{th}}}}{A}\left( {Bx + 1} \right)} \right)}^{\frac{1}{4}}}}}{{\left( {\frac{{{\gamma _{th}}}}{A}\left( {Bx + 1} \right)} \right)}^{\frac{m}{4}}}} {e^{ - \theta {x^{\frac{1}{4}}}}}{x^{\frac{{Nk - 4}}{4}}}dx.
 \vspace{-0.5cm}
\end{align}
\end{small}
\end{spacing}

Furthermore, (\ref{eq:P11}) is evaluated by leveraging the Gauss-Hermite quadrature approach. According to Table (25.10) in \cite{1966Handbook}, one has
\vspace{-0.4cm}
\begin{equation}
\int_{ - \infty }^\infty  {g\left( x \right)} dx = \sum\limits_{w = 1}^W {{\psi  _w}} g\left( {{x_w}} \right)\exp \left( {x_w^2} \right) + {O_W}\,,
\vspace{-0.4cm}
\end{equation}
where $W$ denotes the number of sample points used for approximation, ${{x_w}}$ is the $w$-th root of the Hermite polynomial ${H_W}(x)$~$(w = 1,2, \ldots W)$, ${{\psi  _w}}$ is the $w$-th associated
weight obtained from $\frac{{{2^{W - 1}}W!\sqrt \pi  }}{{{W^2}H_{W - 1}^2({x_w})}}$ and ${O_W}$ is the residual term that tends to $0$ when $W$ tends to infinity.

Using variable substitution $\ln ({x}) = u$ to obtain the new limits with integral from $ - \infty $ to $ + \infty $ and after some mathematical operations, one obtains
\vspace{-0.5cm}
\begin{spacing}{1}
\begin{small}
\begin{align}
\label{eq:P111}
&\int\limits_0^\infty  {F\left( {\frac{{{\gamma _{th}}}}{A}\left( {Bx + 1} \right)} \right)} f\left( x \right)dx\nonumber\\
&  \approx  1 \!\!-\!\! \sum\limits_{m = 0}^{Nk - 1} {\frac{{{\theta ^{Nk + m}}}}{{4m!\left( {Nk - 1} \right)!}}} \sum\limits_{w = 1}^W {{\psi _w}} {e^{ - \theta {{\left( {\frac{{{\gamma _{th}}}}{A}\left( {B{e^{{u_w}}} + 1} \right)} \right)}^{\frac{1}{4}}}}}
 \!\! {\left( {\frac{{{\gamma _{th}}}}{A}\left( {B{e^{{u_w}}} + 1} \right)} \!\!\right)^{\frac{m}{4}}}\!\!{e^{ - \theta {e^{\frac{{{{u_w}}}}{4}}}}}\!\!{e^{{{u_w}}\frac{{Nk - 4}}{4}}}{e^{{u_w} + u_w^2}} \!-\! {O_W}.
 \vspace{-0.5cm}
\end{align}
\end{small}
\end{spacing}

The second integral in (\ref{eq:P1}) can be calculated as
\vspace{-0.3cm}
\begin{spacing}{1}
\begin{small}
\begin{align}
\label{eq:P12}
&\int\limits_{\frac{{{\gamma _{th}}}}{B}}^\infty  {F\left( {\frac{{Bx}}{{A{\gamma _{th}}}} - \frac{1}{A}} \right)} f\left( x \right)dx = \int\limits_{\frac{{{\gamma _{th}}}}{B}}^\infty   \frac{1}{{\Gamma \left( {Nk} \right)}}\gamma \left( {Nk,\theta {{\left( {\frac{{Bx}}{{A{\gamma _{th}}}} - \frac{1}{A}} \right)}^{\frac{1}{4}}}} \right)\frac{{{\theta ^{Nk}}{e^{ - \theta {x^{\frac{1}{4}}}}}{x^{\frac{{Nk - 4}}{4}}}}}{{4\left( {Nk - 1} \right)!}}dx\nonumber\\
& = 1 - F\left( {\frac{{{\gamma _{th}}}}{B}} \right) - \sum\limits_{m = 0}^{Nk - 1} {\frac{{{\theta ^{Nk + m}}}}{{4m!\left( {Nk - 1} \right)!}}} \int\limits_{\frac{{{\gamma _{th}}}}{B}}^\infty  {{e^{ - \theta {{\left( {\frac{{Bx}}{{A{\gamma _{th}}}} - \frac{1}{A}} \right)}^{\frac{1}{4}}}}}} {\left( {\frac{{Bx}}{{A{\gamma _{th}}}} - \frac{1}{A}} \right)^{\frac{m}{4}}}{e^{ - \theta {x^{\frac{1}{4}}}}}{x^{\frac{{Nk - 4}}{4}}}dx.
\vspace{-0.5cm}
\end{align}
\end{small}
\end{spacing}

Similarly, using variable substitution $\ln \left( {x - \frac{{{\gamma _{th}}}}{B}} \right) = u$ to obtain the new limits with integral from $ - \infty $ to $ + \infty $ and after some mathematical operations, one obtains
\vspace{-0.4cm}
\begin{spacing}{1}
\begin{small}
\begin{align}
\label{eq:P122}
&\int\limits_{\frac{{{\gamma _{th}}}}{B}}^\infty  {F\left( {\frac{{Bx}}{{A{\gamma _{th}}}} - \frac{1}{A}} \right)} f\left( x \right)dx \approx 1 - F\left( {\frac{{{\gamma _{th}}}}{B}} \right) - \sum\limits_{m = 0}^{Nk - 1} {\frac{{{\theta ^{Nk + m}}}}{{4m!\left( {Nk - 1} \right)!}}} \sum\limits_{w = 1}^W {{\psi _w}} {e^{ - \theta {{\left( {\frac{{B\left( {{e^{{u_w}}} + \frac{{{\gamma _{th}}}}{B}} \right)}}{{A{\gamma _{th}}}} - \frac{1}{A}} \right)}^{\frac{1}{4}}}}}\nonumber\\
 &\times \!\!{\left(\!\! {\frac{{B\left( {{e^{{u_w}}} + \frac{{{\gamma _{th}}}}{B}} \right)}}{{A{\gamma _{th}}}} - \frac{1}{A}} \!\!\right)^{\frac{m}{4}}}\!\!\!{e^{ - \theta {{\left( {{e^{{u_w}}} + \frac{{{\gamma _{th}}}}{B}} \right)}^{\frac{1}{4}}}}}\!\!{\left( {{e^{{u_w}}} + \frac{{{\gamma _{th}}}}{B}} \right)^{\frac{{Nk - 4}}{4}}}\!\!\!\!{e^{{u_w} + u_w^2}} - {O_W}.
 \vspace{-0.5cm}
\end{align}
\end{small}
\end{spacing}

Then, the second part of (\ref{eq:Poutt}), termed as $P_2$, can be computed as
\vspace{-0.3cm}
\begin{spacing}{1}
\begin{small}
\begin{align}
\label{eq:P2}
{P_2} &= \Pr \left(\! {A{{\left| {{G_{{h_i}{g_{t,i}}}}} \right|}^4} < {\gamma _{th}},\frac{{B{{\left| {{G_{{h_i}{g_{r,i}}}}} \right|}^4}}}{{A{{\left| {{G_{{h_i}{g_{t,i}}}}} \right|}^4} + 1}} \ge {\gamma _{th}}} \!\right)
 = \Pr \left( \!{{{\left| {{G_{{h_i}{g_{t,i}}}}} \right|}^4} < \frac{{{\gamma _{th}}}}{A},{{\left| {{G_{{h_i}{g_{r,i}}}}} \right|}^4} \ge \frac{{{\gamma _{th}}}}{B}\left( \!{A{{\left| {{G_{{h_i}{g_{t,i}}}}} \right|}^4} + 1} \!\right)} \!\right)\nonumber\\
 &\!= \!\!\int\limits_0^{\frac{{{\gamma _{th}}}}{A}}\!\! {1 - F\left( {\frac{{{\gamma _{th}}}}{B}\left( {Ax + 1} \right)} \right)} f\left( x \right)dx
 \!= \!\!\sum\limits_{m = 0}^{Nk - 1} {\frac{{{\theta ^{Nk + m}}}}{{4m!\left( {Nk - 1} \right)!}}}\!\! \int\limits_0^{\frac{{{\gamma _{th}}}}{A}}\!\!\! {{e^{ - \theta {{\left( {\frac{{{\gamma _{th}}}}{B}\left( {Ax + 1} \right)} \right)}^{\frac{1}{4}}}}}{{\left( {\frac{{{\gamma _{th}}}}{B}\left( {Ax + 1} \right)} \right)}^{\frac{m}{4}}}}
{e^{ - \theta {x^{\frac{1}{4}}}}}{x^{\frac{{Nk - 4}}{4}}}dx.
\vspace{-0.5cm}
\end{align}
\end{small}
\end{spacing}
Using (\ref{eq:P111}), (\ref{eq:P122}), and (\ref{eq:P2}), (\ref{eq:Poutt}) can be derived.
\section{}
First, we derive the success probability ${\Phi _{TEP}}$ of the TEP scheme. Since the data of both $U_r$ and $U_t$ have to be successfully decoded, ${\Phi _{TEP}}$ is expressed as
\vspace{-0.3cm}
\begin{spacing}{1}
\begin{small}
\begin{align}
\label{eq:SP}
{\Phi _{TEP}} &\!=\! \Pr \left(\!\! {\frac{{{\gamma _{r,TEP}}}}{{{\gamma _{t,TEP}} + 1}} \!\!>\!\! {\gamma _{th}},{\gamma _{t,TEP}} \!\!>\!\! {\gamma _{th}},{\gamma _{r,EEP}} \!\!\ge\!\! {\gamma _{t,EEP}}} \!\! \right)
 \!+\! \Pr \left( \!\!{\frac{{{\gamma _{t,TEP}}}}{{{\gamma _{r,TEP}} + 1}} \!\!>\!\! {\gamma _{th}},{\gamma _{r,TEP}} \!\!>\!\! {\gamma _{th}},{\gamma _{t,EEP}} \!\!\ge\!\! {\gamma _{r,EEP}}}\! \!\right).
\vspace{-0.3cm}
\end{align}
\end{small}
\end{spacing}

Let ${U_1} = {P_{AP}}l_r^2{\beta _{r,TEP}}{\alpha _r}$, ${U_2} = {P_{AP}}l_t^2{\beta _{t,TEP}}{\alpha _t}$,
the first part of (\ref{eq:SP}), termed as ${\Phi _1}$, is calculated as
\vspace{-0.8cm}
\begin{spacing}{1}
\begin{small}
\begin{align}
&{\Phi _1} \!=\! \Pr \left(\! {{{\left| {{G_{{h_i}{g_{r,i}}}}} \right|}^4} \!>\! \frac{{{\gamma _{th}}({U_2}{{\left| {{G_{{h_i}{g_{t,i}}}}} \right|}^4} + {\alpha _{AP}}{N_0})}}{{{U_1}}},{{\left| {{G_{{h_i}{g_{t,i}}}}} \right|}^4} \!>\! \frac{{{\gamma _{th}}{\alpha _{AP}}{N_0}}}{{{U_2}}},{{\left| {{G_{{h_i}{g_{r,i}}}}} \right|}^4} \!>\! \frac{{{U_2}{{\left| {{G_{{h_i}{g_{t,i}}}}} \right|}^4}}}{{{U_1}}}}\! \right)\nonumber\\
 &= \Pr \left( {{{\left| {{G_{{h_i}{g_{r,i}}}}} \right|}^4} > \frac{{{\gamma _{th}}({U_2}{{\left| {{G_{{h_i}{g_{t,i}}}}} \right|}^4} + {\alpha _{AP}}{N_0})}}{{{U_1}}},{{\left| {{G_{{h_i}{g_{t,i}}}}} \right|}^4} > \frac{{{\gamma _{th}}{\alpha _{AP}}{N_0}}}{{{U_2}}}} \right)\nonumber\\
 &= \int\limits_{\frac{{{\gamma _{th}}{\alpha _{AP}}{N_0}}}{{{U_2}}}}^\infty  {\left( {1 - F\left( {\frac{{{\gamma _{th}}({U_2}x + {\alpha _{AP}}{N_0})}}{{{U_1}}}} \right)} \right)} f\left( x \right)dx\nonumber\\
 &= \int\limits_{\frac{{{\gamma _{th}}{\alpha _{AP}}{N_0}}}{{{U_2}}}}^\infty   \left[ {{e^{ - {{\left( {\frac{{{\theta ^4}{\gamma _{th}}({U_2}x + {\alpha _{AP}}{N_0})}}{{{U_1}}}} \right)}^{\frac{1}{4}}}}}\sum\limits_{m = 0}^{Nk - 1} {\frac{{{{\left( {\frac{{{\theta ^4}{\gamma _{th}}({U_2}x + {\alpha _{AP}}{N_0})}}{{{U_1}}}} \right)}^{\frac{m}{4}}}}}{{m!}}} } \right]\frac{{{\theta ^{Nk}}{e^{ - \theta {x^{\frac{1}{4}}}}}{x^{\frac{{Nk - 4}}{4}}}}}{{4\left( {Nk - 1} \right)!}}dx\nonumber\\
& = \!\sum\limits_{m = 0}^{Nk - 1}\! {\frac{{{\theta ^{Nk}}}}{{4\left( {Nk - 1} \right)!m!}}}\!\! \int\limits_{\frac{{{\gamma _{th}}{\alpha _{AP}}{N_0}}}{{{U_2}}}}^\infty \!\!\!\!\!\!\! {{e^{ - {{\left( {\frac{{{\theta ^4}{\gamma _{th}}({U_2}x + {\alpha _{AP}}{N_0})}}{{{U_1}}}} \right)}^{\frac{1}{4}}}}}} {\left( {\frac{{{\theta ^4}{\gamma _{th}}({U_2}x + {\alpha _{AP}}{N_0})}}{{{U_1}}}} \right)^{\frac{m}{4}}}{e^{ - \theta {x^{\frac{1}{4}}}}}{x^{\frac{{Nk - 4}}{4}}}dx.
\vspace{-0.5cm}
\end{align}
\end{small}
\end{spacing}

Similar to (\ref{eq:P122}),  using variable substitution $\ln \left( {x - \frac{{{\gamma _{th}}{\alpha _{AP}}{N_0}}}{{{U_2}}}} \right) = u$ to obtain the new limits with integral from $ - \infty $ to $ + \infty $ and after some mathematical operations, one obtains
\vspace{-0.3cm}
\begin{small}
\begin{align}
\label{eq:FT1}
{\Phi _1} &= \sum\limits_{m = 0}^{Nk - 1}  \frac{{{\theta ^{Nk}}}}{{4\left( {Nk - 1} \right)!m!}}\sum\limits_{w = 1}^W  {\psi _w}{e^{ - {{\left( {\frac{{{\theta ^4}{\gamma _{th}}{U_3}}}{{{U_1}}}} \right)}^{\frac{1}{4}}}}}
{\left( {\frac{{{\theta ^4}{\gamma _{th}}{U_3}}}{{{U_1}}}} \right)^{\frac{m}{4}}}{e^{ - \theta {{\left( {{e^{{u_w}}} + \frac{{{\gamma _{th}}{\alpha _{AP}}{N_0}}}{{{U_2}}}} \right)}^{\frac{1}{4}}}}}\nonumber\\
& \times {\left( {{e^{{u_w}}} + \frac{{{\gamma _{th}}{\alpha _{AP}}{N_0}}}{{{U_2}}}} \right)^{\frac{{Nk - 4}}{4}}}{e^{{u_w} + u_w^2}} + {O_W},
\vspace{-0.5cm}
\end{align}
\end{small}
where ${U_3} = {U_2}{e^{{u_w}}} + {\gamma _{th}}{\alpha _{AP}}{N_0} + {\alpha _{AP}}{N_0}$.

Using the same method, the second part of (\ref{eq:SP}), termed as ${\Phi _2}$, is calculated as
\vspace{-0.3cm}
\begin{small}
\begin{align}
\label{eq:FT2}
{\Phi _2} &= \sum\limits_{m = 0}^{Nk - 1}  \frac{{{\theta ^{Nk}}}}{{4\left( {Nk - 1} \right)!m!}}\sum\limits_{w = 1}^W  {\psi _w}{e^{ - {{\left( {\frac{{{\theta ^4}{\gamma _{th}}{U_4}}}{{{U_2}}}} \right)}^{\frac{1}{4}}}}}
{\left( {\frac{{{\theta ^4}{\gamma _{th}}{U_4}}}{{{U_2}}}} \right)^{\frac{m}{4}}}{e^{ - \theta {{\left( {{e^{{u_w}}} + \frac{{{\gamma _{th}}{\alpha _{AP}}{N_0}}}{{{U_1}}}} \right)}^{\frac{1}{4}}}}}\nonumber\\
 &\times {\left( {{e^{{u_w}}} + \frac{{{\gamma _{th}}{\alpha _{AP}}{N_0}}}{{{U_1}}}} \right)^{\frac{{Nk - 4}}{4}}}{e^{{u_w} + u_w^2}} + {O_W},
 \vspace{-0.8cm}
\end{align}
\end{small}
where ${U_4} = {U_1}{e^{{u_w}}} + {\gamma _{th}}{\alpha _{AP}}{N_0} + {\alpha _{AP}}{N_0}$.
Using (\ref{eq:FT1}) and (\ref{eq:FT2}), (\ref{eq:TEP}) can be obtained.

The success probability ${\Phi _{T}}$ of the STAR-RIS aided wireless-powered TDMA system is expressed as
\vspace{-1cm}
\begin{spacing}{1}
\begin{small}
\begin{eqnarray}
\label{eq:fT}
{\Phi _{T}} = \Pr \left( {{\gamma _r} > {\gamma _t},{\gamma _t} > {\gamma _{th}}} \right) + \Pr \left( {{\gamma _t} > {\gamma _r},{\gamma _r} > {\gamma _{th}}} \right).
\vspace{-1cm}
\end{eqnarray}
\end{small}
\end{spacing}
The first term of (\ref{eq:fT}), termed as ${\Phi _{T,1}}$, is calculated as
\vspace{-0.3cm}
\begin{small}
\begin{align}
\label{eq:fT1}
&{\Phi _{T,1}} = \Pr \left( \begin{array}{l}
\frac{{{P_{AP}}l_r^2{{\left| {{G_{{h_i}{g_{r,i}}}}} \right|}^4}{\alpha _r}}}{{\alpha _{AP}^r{N_0}}} > \frac{{{P_{AP}}l_t^2{{\left| {{G_{{h_i}{g_{t,i}}}}} \right|}^4}{\alpha _t}}}{{\alpha _{AP}^t{N_0}}},
\frac{{{P_{AP}}l_t^2{{\left| {{G_{{h_i}{g_{t,i}}}}} \right|}^4}{\alpha _t}}}{{\alpha _{AP}^t{N_0}}} > {\gamma _{th}}
\end{array} \right)\nonumber\\
 &= \Pr \left( {\frac{{{\gamma _{th}}\alpha _{AP}^t{N_0}}}{{{P_{AP}}l_t^2{\alpha _t}}} < {{\left| {{G_{{h_i}{g_{t,i}}}}} \right|}^4} < \frac{{l_r^2{{\left| {{G_{{h_i}{g_{r,i}}}}} \right|}^4}{\alpha _r}\alpha _{AP}^t}}{{\alpha _{AP}^rl_t^2{\alpha _t}}}} \right)
 = F\left( {{U_5}{{\left| {{G_{{h_i}{g_{r,i}}}}} \right|}^4}} \right) - F\left( {\frac{{{\gamma _{th}}\alpha _{AP}^t{N_0}}}{{{P_{AP}}l_t^2{\alpha _t}}}} \right),
 \vspace{-0.3cm}
\end{align}
\end{small}
where ${U_5} = \frac{{l_r^2{\alpha _r}\alpha _{AP}^t}}{{\alpha _{AP}^rl_t^2{\alpha _t}}}$.

$F\left( {{U_5}{{\left| {{G_{{h_i}{g_{r,i}}}}} \right|}^4}} \right)$ as the first term of (\ref{eq:fT1}), termed as $\Phi _{T,1}^1$, can be further expressed as
\vspace{-0.5cm}
\begin{spacing}{1}
\begin{small}
\begin{eqnarray}
\label{eq:fT11}
\Phi _{T,1}^1
 = \int\limits_0^\infty  {F\left( {{U_5}x} \right)} f\left( x \right)dx
 = \!\!1\! -\!\! \sum\limits_{m = 0}^{Nk - 1} \!\!{\frac{{{\theta ^{m + Nk}}}}{{m!4\left( {Nk - 1} \right)!}}} \int\limits_0^\infty  {{e^{ - \theta {{\left( {{U_5}x} \right)}^{\frac{1}{4}}}}}} {\left( {{U_5}x} \right)^{\frac{m}{4}}}{e^{ - \theta {x^{\frac{1}{4}}}}}{x^{\frac{{Nk - 4}}{4}}}dx.
 \vspace{-0.5cm}
\end{eqnarray}
\end{small}
\end{spacing}

Let ${x^{\frac{1}{4}}} = t$, (\ref{eq:fT11}) can be written as
\vspace{-0.5cm}
\begin{spacing}{1}
\begin{small}
\begin{eqnarray}
\Phi _{T,1}^1
= 1 - \sum\limits_{m = 0}^{Nk - 1} {\frac{{{\theta ^{m + Nk}}{U_5}^{\frac{m}{4}}}}{{m!\left( {Nk - 1} \right)!}}} \int\limits_0^\infty  {{e^{ - \theta \left( {{U_5}^{\frac{1}{4}} + 1} \right)t}}} {t^{Nk + m - 1}}dt.
\vspace{-0.3cm}
\end{eqnarray}
\end{small}
\end{spacing}

Using (3.478) in \cite{Gradshte1965Table}, one has
\vspace{-0.5cm}
\begin{spacing}{1}
\begin{small}
\begin{eqnarray}
\label{eq:fT111}
\Phi _{T,1}^1
\!= \!\!1 \!\!-\!\! \sum\limits_{m = 0}^{Nk - 1} {\frac{{{\theta ^{m + Nk}}{U_5}^{\frac{m}{4}}}}{{m!\left( {Nk - 1} \right)!}}} {\left( {\theta \left( {{U_5}^{\frac{1}{4}} + 1} \right)} \right)^{ - \left( {Nk + m} \right)}}\!\!\!\!\!\!\!\!\!\!\!\!\Gamma \left( {Nk + m} \right).
\vspace{-0.5cm}
\end{eqnarray}
\end{small}
\end{spacing}

The second part of (\ref{eq:fT}), termed as ${\Phi _{T,2}}$, can be calculated in a similar manner, given by
\vspace{-0.3cm}
\begin{small}
\begin{eqnarray}
\label{eq:fT2}
&{\Phi _{T,2}} = 1 - \sum\limits_{m = 0}^{Nk - 1} {\frac{{{\theta ^{m + Nk}}{U_6}^{\frac{m}{4}}}}{{m!\left( {Nk - 1} \right)!}}} {\left( {\theta \left( {{U_6}^{\frac{1}{4}} + 1} \right)} \right)^{ - \left( {Nk + m} \right)}}
 \Gamma \left( {Nk + m} \right) - F\left( {\frac{{{\gamma _{th}}\alpha _{AP}^r{N_0}}}{{{P_{AP}}l_r^2{\alpha _r}}}} \right),
 \vspace{-0.8cm}
\end{eqnarray}
\end{small}
where ${U_6} = \frac{{l_t^2{\alpha _t}\alpha _{AP}^r}}{{\alpha _{AP}^tl_r^2{\alpha _r}}}$.
Using (\ref{eq:fT1}), (\ref{eq:fT111}), and (\ref{eq:fT2}), (\ref{eq:TDMA}) can be obtained.

\vspace{-0.5cm}


\end{document}